\documentclass{vldb}
\usepackage{times,amsmath,epsfig}

   \usepackage{subfigure,stmaryrd}

   \usepackage{tikz}
   \usetikzlibrary{matrix,arrows}
   \usepackage{multirow}
   \usepackage{alltt}
   \usepackage{color}
   \usepackage{hhline}
   \usepackage{listings}

\newcounter{enum}
\newenvironment{packed_enum}{
\begin{list}{\textbf{(\arabic{enum}.)}}{
  \setlength{\itemsep}{0pt}
  \setlength{\parskip}{0pt}
  \setlength{\labelwidth}{-4 pt}
  \setlength{\leftmargin}{0 pt}
  \setlength{\itemindent}{0pt}
  \usecounter{enum}}
}{\end{list}}

\definecolor{orange}{rgb}{1,0.5,0}
\definecolor{lightpurple}{rgb}{0.5,0,0.5}
\definecolor{purple}{rgb}{0.5,0,0.25}
\definecolor{mybrown}{rgb}{0.5,0,0}

\newcommand{\hide}[1]{}

\usepackage{graphicx}
  \usepackage[pdftex,colorlinks=true,citecolor=black,urlcolor=black,linkcolor=black]{hyperref}

\title{Graph Analytics using the Vertica Relational Database}

\numberofauthors{1}
\author{
\alignauthor
\vspace*{-0.2cm}Alekh Jindal$^\text{\large $\ast$}$\hspace*{0.6cm} Samuel Madden$^\text{\large $\ast$}$\hspace*{0.6cm} Mal\'u Castellanos$^\text{\large $\star$}$\hspace*{0.6cm} Meichun Hsu$^\text{\large $\star$}$\\\vspace*{0.3cm}
      \affaddr{$^\text{\large $\ast$}$CSAIL, MIT\hspace*{2cm} $^\text{\large $\star$}$Vertica, HP Software}
}

\begin{document}
\maketitle
\begin{abstract} 
Graph analytics is becoming increasingly popular, with a deluge of new systems for graph analytics having been proposed in the past few years. These systems often start from the assumption that a new storage or query processing system is needed, 
in spite of graph data being often collected and stored in a relational database in the first place.
In this paper, we study Vertica relational database as a platform for graph analytics. 
We show that vertex-centric graph analysis can be translated to SQL queries, typically involving table scans and joins, and that modern column-oriented databases are very well suited to running such queries.
Specifically, we present an experimental evaluation of the Vertica relational database system on a variety of graph analytics, including iterative analysis, a combination of graph and relational analyses, and more complex 1-hop neighborhood graph analytics, showing that it is competitive to two popular vertex-centric graph analytics systems, namely Giraph and GraphLab.
\end{abstract}

\section{Introduction}
\label{section:introduction}

Recent years have seen growing interest in the area of graph data management from businesses and academia. This focus on graphs arises from their use in a number of new applications, including social network analytics, transportation, ad and e-commerce recommendation systems, and web search.  As a result, a deluge of new graph data management systems have been proposed. In particular, many systems focus on {\it graph analytics}, e.g., efficiently computing statistics and other metrics over graphs, such as PageRank or shortest paths. These graph analytics workloads are seen as quite different from traditional database analytics, largely due to the iterative nature of many of these computations, and the perceived awkwardness of expressing graph analytics as SQL queries (which typically involves multiple self-joins on tables of nodes and edges). Examples of these new systems include a number of so-called ``vertex-centric'' systems (e.g., Pregel~\cite{pregel}, Giraph~\cite{giraph}, GraphLab~\cite{graphlab}, Trinity~\cite{trinity}, and Pregelix~\cite{pregelix}).

\subsection{Why Relational Databases?}

Given the popular demand for graph analytics, a natural
question is whether or not traditional database systems really are a bad fit for these graph analytics workloads? This question arises because, in many real-world scenarios, graph data is collected and stored in a relational database in the first place and it is expensive to move data around. Given this, if it is avoidable, users may prefer not to export their data from the relational database into a graph database. Rather, they would like to perform the graph analytics (with comparable performance) directly with the relational engine, without the expensive step of copying data into a file
system (or distributed storage system like HDFS), in order to be processed by a graph system, and then (possibly) back into the relational system for further processing.
Indeed, some early efforts to implement graph queries in relational databases~\cite{oracle-graph,vertica-triangles,graphiso} have shown promise in this regard, but have typically
only evaluated one or a small number of benchmarks, and not demonstrated the feasibility of implementing an efficient, general-purpose graph engine in a relational system.

Apart from the need to avoid copying data in and out of file systems, graph engines suffer from another limitation. As graphs get larger and larger, frequently the users want to (or will have to) select a subset of a graph before performing analysis on it. For example, it is unlikely that a user will run a single-source shortest path query on the entire trillion node Facebook graph --- this would be prohibitively slow on any system.  Rather, it is more likely that users will run several shortest paths queries over different subsets of the graph (e.g., the N-hop neighbors of some particular user.) Furthermore, real-world graphs have vertices and edges accompanied by several other attributes. For example, edges in a social network may be of different types such as friends, family, or classmates. Similarly nodes may have several attributes to describe the properties of each person in the social network, e.g., their username, birthdate, and so on. 
Given such metadata, an analyst would typically do some ad-hoc relational analysis in addition to the graph analysis. For instance, the analyst may want to preprocess and shape the graph before running the actual graph algorithm, e.g.,~filtering edges based on timestamp or limiting the graph to just close friends of a particular user. Similarly, he may want to analyze the output of the graph analysis, e.g.,~computing aggregates to count the number of edges or nodes satisfying some property, or other statistics.
Such pre- and post- processing of graph data   requires relational operators such as selection, projection, aggregation, and join, for which relational databases are highly optimized.

In addition to combining relational analysis, several graph analyses compute aggregates over a larger neighborhood. 
For example, counting the triangles
in a graph requires every vertex to access its neighbors' neighbors (which could potentially form the three vertices of the triangle).
Likewise, finding whether a vertex acts as a bridge (weak ties) between two disconnected vertices requires every vertex to check for the presence of edges between its neighbors, i.e.~the 1-hop neighborhood. 
Vertex-centric interfaces like Pregel~\cite{pregel} are tedious for expressing the above queries, as they require sending neighborhood information to all its neighbors in the first superstep and then performing the actual analysis in the second superstep. 
SQL, on the other hand, is a much more powerful and general purpose language for capturing such analyses. 
For example, in Vertica, we can express triangle counting as a three-way self-join over the edge table very efficiently~\cite{vertica-triangles}. 
Similarly, we can detect weak ties using two inner joins (to get the two vertices on either side of the bridge) and one outer join (to make sure that the two vertices are not connected by an edge). 
Thus, by simply adding more joins, SQL is more flexible at expressing such graph analyses.

Finally, the query optimizer in a relational database picks the best plan for the multi-join queries and the system can take care of re-segmenting the data when needed.
As an example, the optimizer may decide to fully pipeline the triangle counting query and never materialize the intermediate output.
This is in contrast to static query execution plans in typical graph analytics systems such as Giraph.
Furthermore, since graph manipulations in Giraph are not implemented as operators, it is difficult to modify or extend the Giraph execution pipeline.
Relational databases, on the other hand, are extensible by design.

\subsection{Why Column Stores?}

Graph analytics typically involves  scan-oriented table joins followed by aggregates, which are highly suited for column-oriented relational databases like Vertica.
In an earlier effort, we compared the performance of different relational data stores over graph queries and column stores were a clear winner~\cite{istcblog-bench}.
This is due to combination of several features in modern column stores, including efficient data storage (vertical partitioning and compression), vectorized data access, query pipelining, late materialization, and numerous join optimizations.
Furthermore, in contrast to the narrow tables in raw graphs, the presence of metadata results in wide vertex and edge tables, where column stores will perform especially well with as they only need to access  the columns that are relevant for the analysis.

In this paper, we describe four key aspects necessary to build high-performance graph analytics in the Vertica column-oriented database.
First, we look at how we can translate the logical query plans of  vertex-centric graph queries into relational operators and run them as standard SQL. Although vertex compute functions can be rewritten into pure SQL in some cases, we find that table UDFs (offered by many relational databases, including Vertica) are sufficient to express arbitrarily complex vertex functions as well.
Second, we show several query optimization techniques to tune the performance of graph queries on Vertica. These include considering updating vs replacing the nodes table on each iteration, incremental evaluation of queries, and eliminating redundant joins.  
Third, we outline several features specific to a column-store like Vertica that makes it well suited to run graph analytics queries.
Finally, we show how Vertica can be optimized using table UDFs to run iterative graph analytics in-memory, which significantly reduces the disk I/Os (and overall query time) at the cost of higher memory footprint.

\vspace{0.1cm}
\noindent\textbf{Contributions.} In summary, our key contributions are as follows:

\vspace{-0.2cm}
\begin{packed_enum}

\item We take a closer look at vertex-centric graph processing, using the Giraph system (a popular graph analytics system) as an example.
We show that vertex-centric graph processing can be expressed as a query execution plan, which in the case of Giraph is a fixed plan that is used to run all Giraph programs.
We then show that this plan can be expressed as a logical query plan that can be optimized using a relational query optimizer (Section~\ref{section:giraphanalysis}).

\item We show how we can translate this vertex-centric plan into SQL, which can  be run on standard relational databases. We describe several query optimizations to improve the performance of vertex-centric queries and describe Vertica specific features to run these queries efficiently. As a concrete example, we discuss the physical query execution plan of single source shortest path on Vertica. Lastly, we show how Vertica can be extended via table UDFs to run the entire unmodified vertex-centric query in-memory and as a single transaction (Section~\ref{section:vertexcentric}).

\item We provide an extensive experimental evaluation of several typical graph queries on large, billion-edge graphs in Vertica.  We compare it with two popular vertex-centric graph processing systems,  GraphLab and Giraph. Our key findings are: 
(i)~Vertica has comparable end-to-end performance to these popular vertex-centric systems,
(ii)~Vertica has a much smaller memory footprint than other systems, at the cost of much greater disk I/O,
(iii)~We can extend Vertica to trade an increased memory footprint for faster runtimes, comparable to that of GraphLab,
(iv)~relational engines naturally excel at combining graph analysis with relational analysis, and
(v)~relational engines can implement more complex 1-hop neighborhood graph analyses, which vertex-centric programming cannot express efficiently (Section~\ref{section:benchmarks}).

\end{packed_enum}

\section{Background}
\label{section:giraphanalysis}

Vertex-centric graph processing, first proposed in Pregel~\cite{pregel}, has become the most popular general purpose way of 
processing graph data, due to its ease-of-use and proven ability of engines based on it to scale to large graphs~\cite{graphlab,gps,trinity,grace,pregelix}.
In this section, we first recap the vertex-centric programming model. 
Then, to understand the graph processing in a typical vertex-centric system, we analyze the execution pipeline in Giraph, an open-source implementation of Pregel, and express it as a logical query plan.
Other Pregel-like systems use a similar static query plan, though some may use different scheduling strategies, e.g.~GraphLab.

\subsection{Vertex-centric Model}

In the vertex-centric programming model, the user provides a UDF (the \textit{vertex program}) specifying the computation that happens at each vertex of the graph. The UDFs update the vertex state and communicate by sharing messages with neighboring vertices.
The underlying execution engine may choose to run the vertex-centric programs synchronously, as a series of {\it supersteps} with synchronization between then,
 or asynchronously, where threads update a representation of the graph in shared memory.
Programmers do not have to worry about details such how the graph is partitioned across nodes/threads, how it is distributed across multiple machines, or how message passing and coordination works. Each vertex may be on the same physical machine or a different, remote machine. The concept is similar to MapReduce, where programmers only specify map and the reduce functions without worrying about the system details.
To illustrate, Listing~\ref{listing:shortestpath_pregel} shows how a programmer would implement single source shortest paths (SSSP) using Giraph (other Pregel-like systems have very similar syntax).
In this program, each vertex compares its current shortest distance to the source to the distance reported by each of its neighbors, and if a shorter distance is found, updates its distance and propagates the updated distance to its neighbors.

\begin{lstlisting}[basicstyle=\scriptsize,aboveskip=0pt,belowskip=-14pt,float=!t,language=java,label=listing:shortestpath_pregel,caption=Single Source Shortest Path in Giraph.]
public void compute(Iterable<IntWritable> messages) {

  // get the minimum distance
  if (getSuperstep() == 0)
    setValue(new DoubleWritable(Integer.MAX_VALUE));
  int minDist = isSource() ? 0 : Integer.MAX_VALUE;
  for (IntWritable message : messages) 
    minDist = Math.min(minDist, message.get());
    
  // send messages to all edges if new minimum is found 		
  if (minDist < getValue().get()) {
    setValue(new IntWritable(minDist));
    for (Edge<?, ?> edge : getEdges()) {
      int distance = minDist + edge.getValue().get();
      sendMessage(edge.getTargetVertexId(), new IntWritable(distance));
    }
  }
  voteToHalt();	// halt
}
\end{lstlisting}

\subsection{Giraph Execution Pipeline}

We now  provide a detailed study of execution workflow used in Giraph, to illustrate the key steps in vertex-centric program execution.
Giraph runs graph analyses as user provided vertex-centric programs on top of Hadoop MapReduce.
The user  provides the computation that happens at each vertex of the graph and Giraph takes care of running it in a distributed fashion over a large cluster.

\begin{figure}[!t]
\centering
\vspace{-0.1cm}
\includegraphics[width=3.2in]{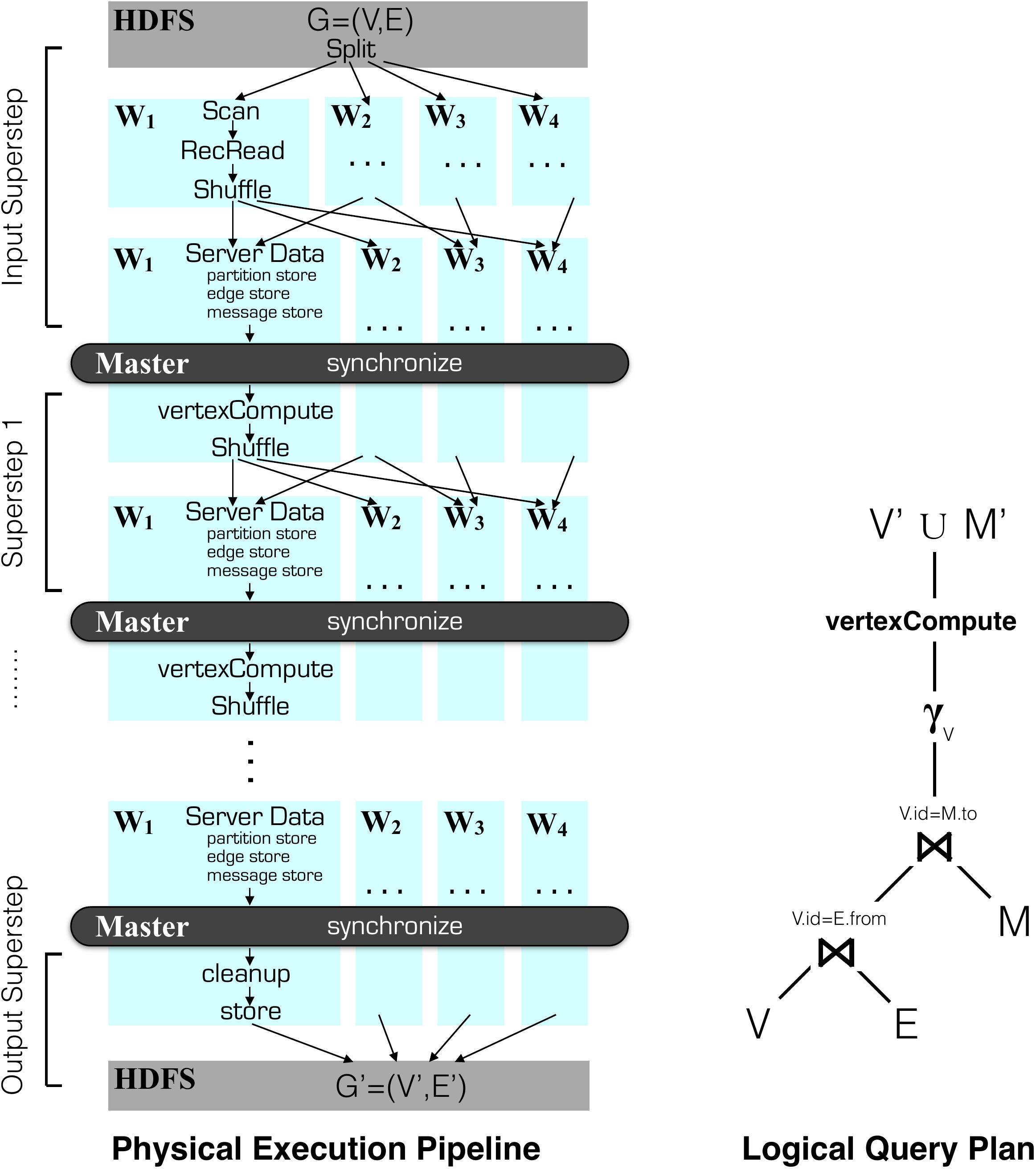}
\vspace{-0.3cm}
\caption{Giraph Physical Execution Pipeline and its Logical Representation.}
\vspace{-0.6cm}
\label{fig:giraphplan}
\end{figure}

Specifically, Giraph executes the vertex-centric query as a map-only job (the \textit{GiraphMapper}) on Hadoop MapReduce.
However, it uses the Hadoop MapReduce infrastructure only to allocate  nodes, using  mappers  simply as containers for Giraph workers.  These map jobs run for the duration of the job, repeatedly executing the compute UDF and communicating with other mappers over sockets.
To illustrate this data flow, the left-side of Figure~\ref{fig:giraphplan} shows the physical execution of Giraph with four workers $W_1$ to $W_4$.
The execution in Giraph is organized into \textit{supersteps}, wherein each worker operates in parallel during the superstep and the workers synchronize at the end of the superstep.
During the \textit{InputSuperstep}, the system splits the input graph into a list of vertices $V$ and list of edges $E$ as it reads data from HDFS.  
Each worker reads the split assigned to it, parses it into vertices and edges, and partitions them across all workers, typically using a hash-based partitioner. 
Each worker then builds its \textit{ServerData}, consisting of three components: (i) the ~\textit{partition store} to keep the partition vertices and related metadata, (ii) the~\textit{edge store} to keep the partition edges and related metadata, and (iii) the~\textit{message store} to keep the incoming messages for this partition. At the end of the InputSuperstep, i.e.~when all workers have finished creating the ServerData, the workers are ready to perform the actual vertex computation.
In each superstep after the InputSuperstep, the workers run the \textit{vertexCompute} UDF for the vertices in their respective partition and shuffle the outgoing messages across all workers. The workers then update their respective ServerData and wait for everyone to finish the superstep (the \textit{synchronize} barrier). 
Finally, when there are no more messages to process, the workers store the output graph back in HDFS during the \textit{OutputSuperstep}. Thus, we see that similar to MapReduce execution in Hadoop~\cite{hadooppp}, Giraph has a static hard-coded query execution pipeline.

\subsection{Logical Query Plan}

The above Giraph physical execution pipeline can also be represented as a logical query plan, consisting of relational operators and the vertexCompute UDF.
The right-side of Figure~\ref{fig:giraphplan} shows such a simplified logical query plan. 
Assuming that the graph structure itself remains unchanged\footnote{This is true for several typical vertex-centric graph analysis, such as PageRank, shortest paths, connected components, etc.}, 
the Giraph execution pipeline is essentially a distributed vertex update query. That is, it takes the set of vertices $V$ (each having an id and a value), edges $E$ (each having a source and a destination vertex id) and messages $M$ (each having destination vertex id and the message value), runs the vertexCompute UDF for each vertex, and produces the set of output vertices ($V'$) and messages ($M'$), as shown on the right in Figure~\ref{fig:giraphplan}.

The downside of the above vertex-centric query execution in Giraph is that all graph analysis is forced to fit into a fixed query plan.
This is not desirable for several analyses. For example, triangle counting, which requires a three-way join over the edges table, is very difficult to fit in this model.
Moreover, the Giraph logical query plan is not really implemented as a composition of query operators, making it very difficult to modify, extend, or add functionality to the execution pipeline. For instance, the join with $M$ is implemented as a sort merge join; changing to another join implementation would require several deep changes in the system.
Furthermore, even if one could  extend or modify the physical execution pipeline, e.g.~switch merge join to hash join, Giraph cannot make dynamic decisions regarding the best physical plan. For example, hash join may be suitable for very large numbers of intermediate messages and merge join better for small numbers of messages. Giraph does not have this flexibility.
Finally, Giraph is a custom built query processor restricted to a specific type of graph analysis. It cannot be used for more broader types of queries, e.g.~multi-hop analysis, or end-to-end graph analysis, e.g.~analyzing the output of graph analysis, or combining multiple graph analyses. 

In the rest of the paper, we show how relational databases can overcome many of these limitations and, in particular, how Vertica is highly suited for a variety of graph analytics.


\section{Graph Analytics using Vertica}
\label{section:vertexcentric}


In this section, we describe how vertex-centric queries can be run in SQL on a relational database like Vertica.
The goal of this section is to show how we can: (i)~translate vertex-centric graph analyses to standard SQL queries, (ii)~apply several query optimizations to improve the performance of graph analyses, and (iii)~leverage key features of Vertica for efficiently executing these graph analytics queries.


\subsection{Translation to SQL}
\label{section:sqltranslation}

\begin{figure}[!t]
\hspace{-0.2cm}
\includegraphics[width=3.6in]{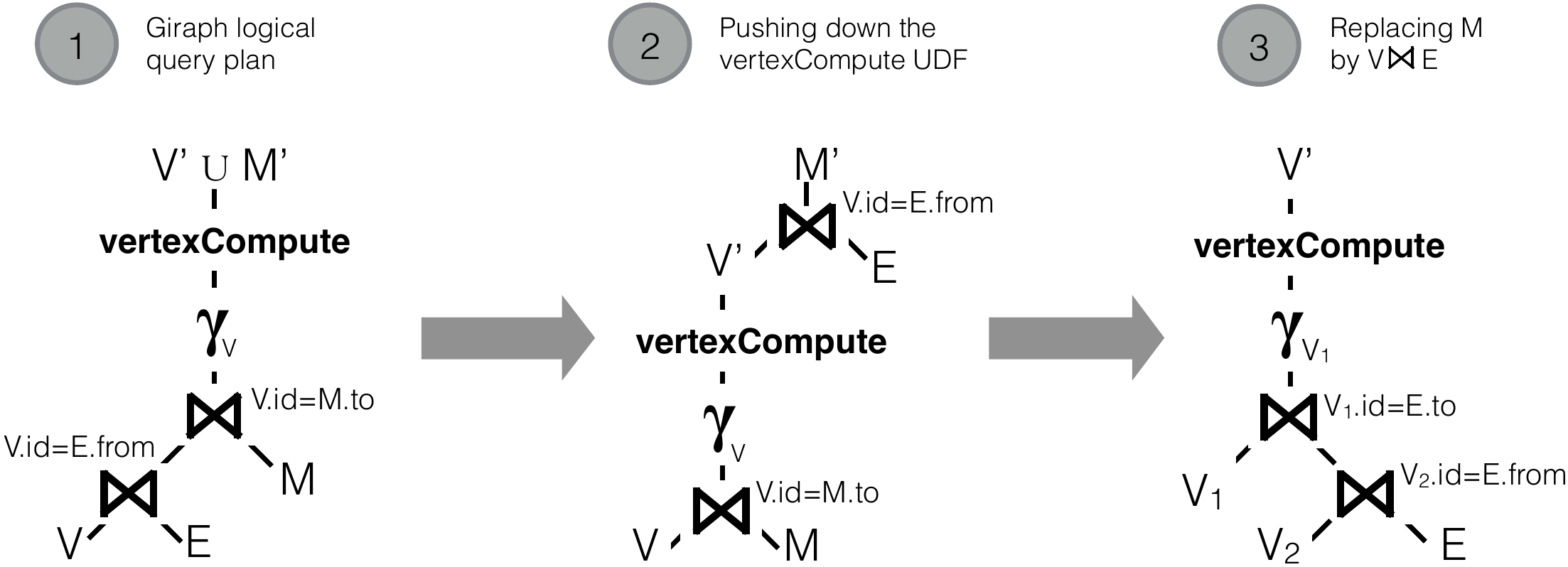}
\vspace{-0.4cm}
\caption{Rewriting Giraph Logical Query Plan.}
\vspace{-0.4cm}
\label{fig:giraphplanrewrite}
\end{figure}

In the following, we describe how we rewrite and translate the Giraph logical query plan to standard SQL.

\subsubsection{Eliminating the message table}

Consider again the Giraph logical query plan, shown on the left in Figure~\ref{fig:giraphplanrewrite}. 
The relation $M$ in this plan is an intermediate output and an artifact of message passing, a consistency mechanism in Giraph.
Since relational databases take care of consistency and allow us to operate directly on the relational tables, we can get rid of $M$. Note that the relation $M$ is used to communicate the new values of vertices to its neighbors. Therefore, we can push down the vertex compute function and obtain the new messages $M'$ by joining the new vertex values ($V'$) with the outgoing edges ($E$), as shown in the middle of Figure~\ref{fig:giraphplanrewrite}. Finally, we can replace $M$ with $V\Join E$ and get rid of relation $M$ completely, as shown on the right in Figure~\ref{fig:giraphplanrewrite}. This simplified query plan deals only with relations $V$ and $E$ as the input and produces modified relation $V'$ as output.


\subsubsection{Translating the vertex compute functions}

\begin{figure}[!t]
\centering
\subfigure[SSSP]{
\includegraphics[width=0.7in]{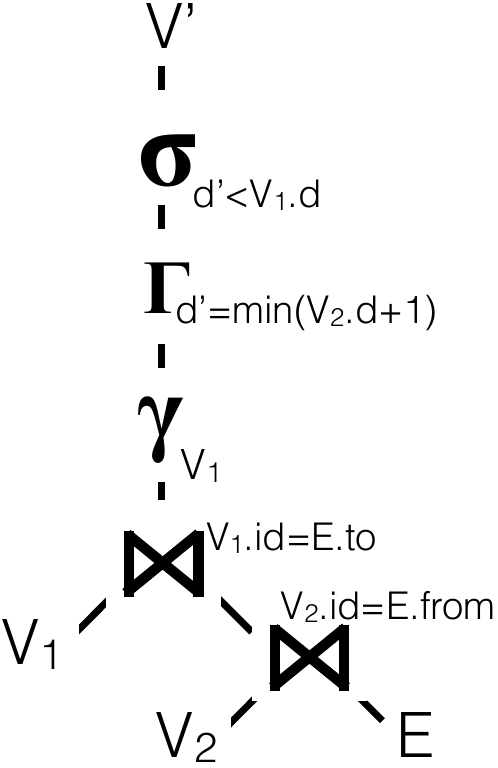}
\label{fig:giraphsssp}
}
\hspace{0.3cm}
\subfigure[CC]{
\includegraphics[width=0.7in]{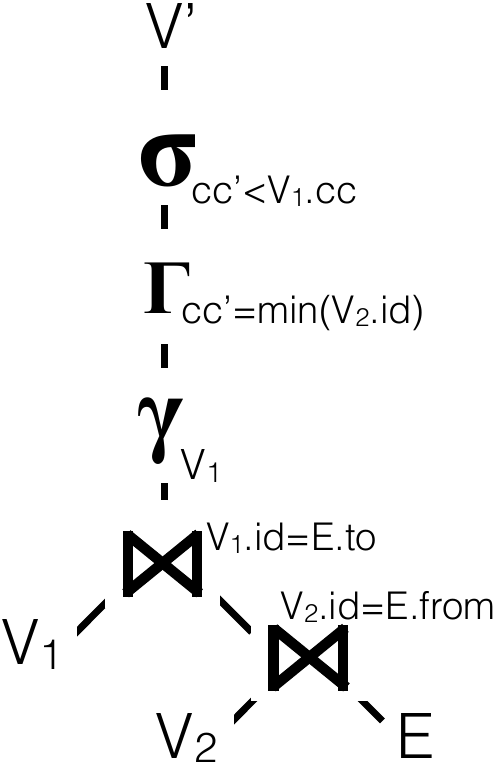}
\label{fig:giraphcc}
}
\hspace{0.3cm}
\subfigure[PageRank]{
\includegraphics[width=0.7in]{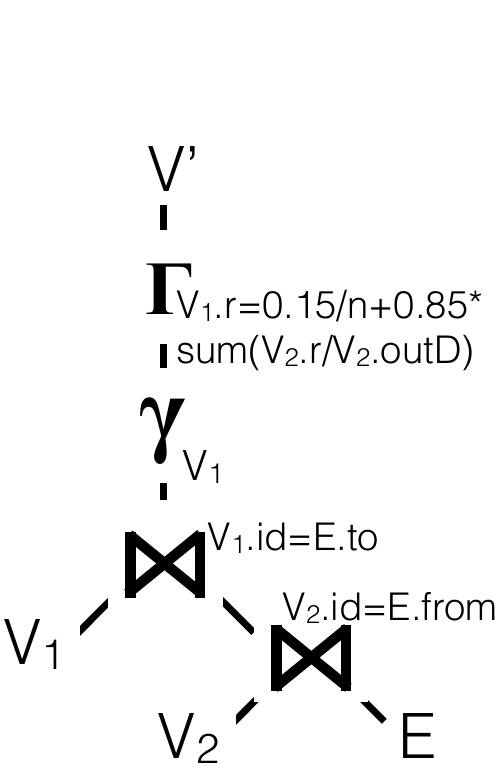}
\label{fig:giraphpr}
}
\vspace{-0.1cm}
\caption{Logical query plans for three vertex-centric queries: (i)~single source shortest path (SSSP), (ii)~connect components (CC), and PageRank.}
\vspace{-0.2cm}
\end{figure}

The \textit{vertexCompute} in the Giraph logical query plan (Figure~\ref{fig:giraphplan}) can be an arbitrary user defined function, similar to \textit{map}/\textit{reduce} in the MapReduce framework.
However, for many graph analytics, the vertex function  involves relatively simple and well defined aggregate operations, which can be expressed directly in relational algebra/SQL. For example, the vertex function for SSSP in Listing~\ref{listing:shortestpath_pregel} finds the MIN of the neighboring distances and applies the filter for detecting smaller distances, i.e.:

\vspace*{-0.4cm}
\begin{equation*}
\text{SSSP}: \textit{\small vertexCompute} \longmapsto \sigma_{\text{d'}<V_1.d} (\Gamma_{\text{d'=min}(V_2.d+1)})
\end{equation*}
\vspace*{-0.4cm}

\noindent Figure~\ref{fig:giraphsssp} shows the resulting logical query plan. Similarly, the vertex function for connected components finds the minimum vertex ID amongst its neighbors and filters for new minimum found (Figures~\ref{fig:giraphcc}), whereas the vertex function for PageRank combines the PageRank of its neighbors (Figure~\ref{fig:giraphpr}), i.e.:

\vspace*{-0.3cm}
\begin{equation*}
\text{CC}: \textit{\small vertexCompute} \longmapsto \sigma_{\text{cc'}<V_1.cc} (\Gamma_{\text{cc'=min}(V_2.id)})
\end{equation*}
\vspace*{-0.5cm}

\vspace*{-0.5cm}
\begin{equation*}
\text{PageRank}: \textit{\small vertexCompute} \longmapsto \Gamma_{V_1.r = \frac{0.15}{n} + 0.85*sum(\frac{V_2.r}{V_2.outD})}
\end{equation*}
\vspace*{-0.3cm}

\noindent By rewriting the vertex functions as relational expressions, the resulting query plans become purely relational and can be implemented completely in standard SQL, without using
user-defined function features in the database system at all (we describe a more general implementation based on UDFs in Section~\ref{sec:udfs} below).
For instance, we could write SSSP from Figure~\ref{fig:giraphsssp} as the following SQL expression:

\begin{scriptsize}
\begin{alltt}
SELECT v1.id, MIN(v2.d+1) AS d
    FROM vertex AS v1, edge AS e, vertex AS v2
    WHERE v2.id = e.from_node AND v1.id = e.to_node
    GROUP BY e.to_node, v1.d
    HAVING MIN(v2.d+1) < v1.d
\end{alltt}
\end{scriptsize}

\noindent The above SQL query computes the minimum neighboring distance of every vertex and filters, via the HAVING clause, distances that are smaller than the already known distance. The resulting vertices can then be used to update the \texttt{\small vertex} relation, as shown in Listing~\ref{listing:shortestpath_vertica}.

\begin{lstlisting}[basicstyle=\scriptsize,aboveskip=0pt,belowskip=-8pt,float=!h,language=sql,label=listing:shortestpath_vertica,caption=Shortest Path in SQL.]
UPDATE vertex AS v SET v.d=v'.d 
  FROM (
    SELECT v1.id, MIN(v2.d+1) AS d
      FROM vertex AS v1, edge AS e, vertex AS v2
      WHERE v2.id = e.from_node AND v1.id = e.to_node
      GROUP BY e.to_node, v1.d
      HAVING MIN(v2.d+1) < v1.d
    ) AS v'
    WHERE v.id=v'.id;
\end{lstlisting}


Finally, a driver program (run via a stored procedure) repeatedly runs the above shortest path query as long as there are any updates, i.e.~at least one of the vertices finds a shorter distance.

\subsection{Query Optimizations}
\label{section:sql_optimizations}

The advantage of expressing graph analysis as relational queries is that we can apply several relational query optimizations, i.e.~we have the flexibility to optimize the queries in several different ways in order to boost performance, in contrast to the hard-coded execution pipeline in Giraph. 
In the following, we present three such query optimizations that can be used to tune the performance. 
We use vertex-centric single source shortest paths as the running example. 
However, of course, these optimizations are applicable in general to SQL-based graph analytics.

\subsubsection{Update Vs Replace}
\label{ref:opt_Update_Vs_Replace}

Graph queries often involve updating large portions of the graph over and over again. However, large number of updates can be a barrier to good performance, especially in read optimized systems like Vertica. To overcome this problem, we can instead replace the vertex or edge table with a new copy of tables containing the updated values. For instance, the single source shortest path involves updating all vertices that find a smaller distance in an iteration. As we explore the graph in parallel, the number of such vertices can quickly grow very large. Therefore, instead of updating the vertices in the existing vertex relation, we can create a new vertex relation (\texttt{\small vertex\_prime}) by joining the updated vertices with the non-updated vertices:

\begin{lstlisting}[basicstyle=\scriptsize,aboveskip=0pt,belowskip=-8pt,float=!h,language=sql,label=listing:opt_updatevsreplace,caption=Shortest Path with Replace instead of Update.]
CREATE TABLE vertex_prime AS
  SELECT v.id, ISNULL(v'.d, v.d) AS d
    FROM vertex AS v LEFT JOIN (
      SELECT v1.id AS id, MIN(v2.d+1) AS d
        FROM vertex AS v1, edge AS e, vertex AS v2
        WHERE v2.id=e.from_node AND v1.id=e.to_node
        GROUP BY e.to_node, v1.d
        HAVING MIN(v2.d+1) < v1.d
    ) AS v'
    ON v.id = v'.Id;
\end{lstlisting}

\noindent Afterwards, we replace \texttt{\small vertex} with \texttt{\small vertex\_prime}. This replacement is quite fast, and, in general, creating a new table is faster than updating because it allows new records to be written sequentially to the table, rather than performing random I/O to update-in-place or recording large delete lists in Vertica.  One downside of this approach is that we lose the physical design (i.e., indexes) on the original table, and physical designs are expensive to create during query execution. 
However, many graph analyses, including PageRank and SSSP, update only the smaller vertex table and therefore the physical designs on the larger edge table can be preserved.
Still, update-in-place may  be more efficient for algorithms that perform small numbers of updates. For instance, the parallel graph exploration in single source shortest path updates very few vertices in the first few iterations. Therefore, a more sophisticated approach is to apply \textit{updates} in the first few iterations before switching to \textit{replace}. In this work, we experimentally determine a fixed threshold to switch from \textit{updates} to \textit{replace}.  Eventually, of course, a cost-based optimizer could be use to determine when to switch.

\subsubsection{Incremental Evaluation}
\label{ref:opt_Incremental_Evaluation}

Typically, iterative queries process different portions of the data in different iterations.
As a result, there is an opportunity for incremental query evaluation. This is applicable to iterative graph queries as well. For example, in single source shortest path, we do not need to explore the entire graph in every iteration. We need to only explore the neighbors of vertices that found a smaller distance in the previous iteration. This introduces the overhead of keeping track of such vertices from previous iteration, but allows us to benefit by only joining the incrementally updated vertices table (\texttt{\small v\_update}) with its neighbors. To achieve this, we initialize \texttt{\small v\_update} with the \texttt{\small startNode} since that is the only vertex that updated its distance to $0$. Thereafter, in each iteration, we get the new set of updated vertices (\texttt{\small v\_update\_prime}) from the existing set (\texttt{\small v\_update}).
Although we need to materialize additional intermediate output,  we are able to exploit it to significantly reduce the join cardinalities by expanding only the neighbors of the updated vertices. We can then replace \texttt{\small v\_update} by \texttt{\small v\_update\_prime} and get the updated set of vertices.
Listing~\ref{listing:opt_incremental} shows the incrementally evaluated single source shortest path query.  Note that Giraph actually employs a similar optimization as it only computes
updates for active vertices in each superstep.

%
%
%
%
%
%

\begin{lstlisting}[basicstyle=\scriptsize,aboveskip=0pt,belowskip=-4pt,float=!h,language=sql,label=listing:opt_incremental,caption=Shortest Path with Incremental Evaluation.]
CREATE TABLE v_update_prime AS
  SELECT v1.id, MIN(v2.d+1) AS d
    FROM v_update AS v2, edge AS e, vertex AS v1
    WHERE v2.id=e.from_node AND v1.id=e.to_node
    GROUP BY e.to_node, v1.d
    HAVING MIN(v2.d+1) < v1.d;

DROP TABLE v_update; 
ALTER TABLE v_update_prime RENAME TO v_update ;

CREATE TABLE vertex_prime AS
  SELECT v.id, ISNULL(v_update.d, v.d) AS value
    FROM vertex AS v LEFT JOIN v_update
    ON v.id = v_update.id;

DROP TABLE vertex; ALTER TABLE vertex_prime RENAME TO vertex;     
\end{lstlisting}



\subsubsection{Join Elimination}

\begin{figure}[!h]
\centering
\vspace{-0.1cm}
\includegraphics[width=2in]{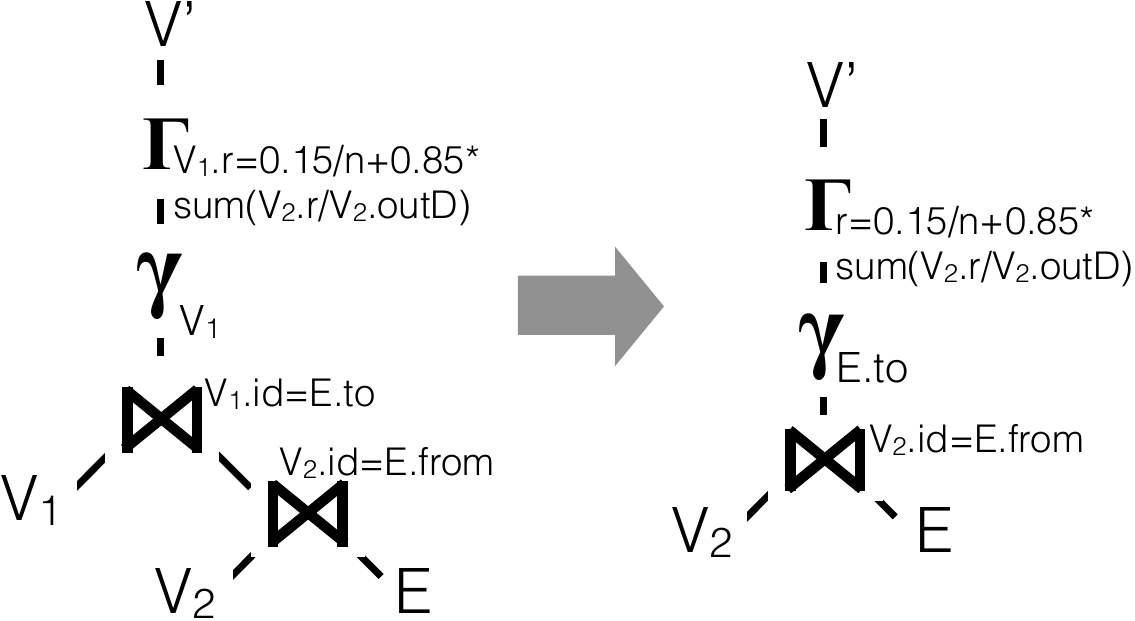}
\vspace{-0.1cm}
\caption{Join Elimination in PageRank.}
\vspace{-0.1cm}
\label{fig:giraph-querypr_joinelimination}
\end{figure}

Several graph analysis perform neighborhood access without reading the metadata associated with the neighboring vertices. This means that even though the logical query plan may have a join between the vertex and the edge table, we read only the vertex id from the vertex table. Thus, the join is redundant and can be eliminated. For example, in the logical query plan for PageRank in Figure~\ref{fig:giraphpr}, we read only the vertex id from $V_1$. Therefore, the join with $V_1$ is redundant and can be eliminated as shown in Figure~\ref{fig:giraph-querypr_joinelimination}.

\noindent Eliminating one of the joins in the above query will result in much better performance because the output of $V_2\Join E$, which is as big as the number of edges itself,  does not need to be re-partitioned again to perform the second join.


%

\vspace{0.05cm}

In summary,  vertex-centric graph analyses can be translated to SQL, enabling several optimization techniques to tune the performance.
In the following section, we look at the actual query execution of these graph queries and describe what makes Vertica a good choice for such analyses.


%
%
%

\subsection{Query Execution}

In the previous section, we saw that graph analyses typically involve full table scans and joins over the vertex and the edge tables. 
We now look at some of the features that makes Vertica well-suited for executing such queries.
Specifically, we describe four key features that Vertica provides: (i)~an optimized physical database design, (ii)~join optimizations, (iii)~query pipelining, and (iv)~intra-query parallelism.
Thereafter, we walk through the query execution plan of single source shortest path in Vertica and contrast it with that of Giraph.


\subsubsection{Physical Design}
\label{ref:opt_Physical_Design}

Vertica provides rich support for creating physical designs in order to boost query performance. For instance, it allows creating projections, sort orders, and segmentations within and across different nodes, as well as several encoding and compression schemes. See~\cite{vertica-phydes} for more details on physical design using Vertica. Although the columnar data representation is not useful for projections over narrow vertex and edge tables with just a few columns, it is useful for efficiently compressing these tables and saving disk I/O. As a result, we can create multiple table projections over these tables, in order to boost the performance of queries, while still not exceeding the raw data size. For instance, we can create two projections over the edge table, one segmented on \texttt{\small from\_node} and the other on \texttt{\small to\_node}, in order to perform different self-joins over the edge table locally. Likewise, we can sort the projections on different attributes for performing merge join instead of hash join as well as for evaluating selection predicates. 

Thus, using Vertica, developers can efficiently encode and compress their graph data, create multiple sort orders and partitionings, and based on the physical design, leverage the query optimizer to automatically pick the best physical query operators for their analysis at run time.

%

\subsubsection{Join Optimizations}
\label{ref:opt_Join_Optimizations}

Graph analyses when written in SQL make heavy use of joins. Vertica is highly optimized to efficiently execute such joins over large tables. For example, it can perform joins directly on compressed data without decoding it, apply type dependent just-in-time compilation of the join condition in order to avoid branching, and use sideways information passing (SIP) to push down the join condition as selection predicate over the outer input and thus filter tuples early on~\cite{vertica7years}. Furthermore, databases are not limited to a specific join implementation. Rather, the optimizer can choose between hash or merge joins, or even dynamically switch between the two.

Efficient join processing is a key feature that makes  graph analysis possible in Vertica, by allowing developers to quickly traverse and manipulate large graphs via repeated self-joins.

\subsubsection{Query Pipelining}
\label{ref:opt_Query_Pipelining}

%

Vertica supports pipelined query execution, which avoids materializing intermediate results that would otherwise require repeated access to disk. This is important because graph queries involve join operations that can have large intermediate results which can benefit dramatically from pipelining. For instance, in each iteration, the single source shortest path joins a vertex with its incoming edges and incoming nodes, thereby resulting in an intermediate result with cardinality equal to the number of edges.
We can induce pipelining for such queries by creating sort orders on join and group by attributes. Additionally, we can express graph operations as nested queries, allowing the query optimizer to employ pipelining between the inner and outer query when possible. This is in contrast to Giraph, which blocks the execution and materializes all intermediate output before running the vertex compute function.

Thus, pipelining allows Vertica to avoid materializing large intermediate outputs, which are typical in graph queries.  This reduces memory footprint and improves performance.

%
%

\begin{figure}[!t]
\centering
\includegraphics[width=3.5in]{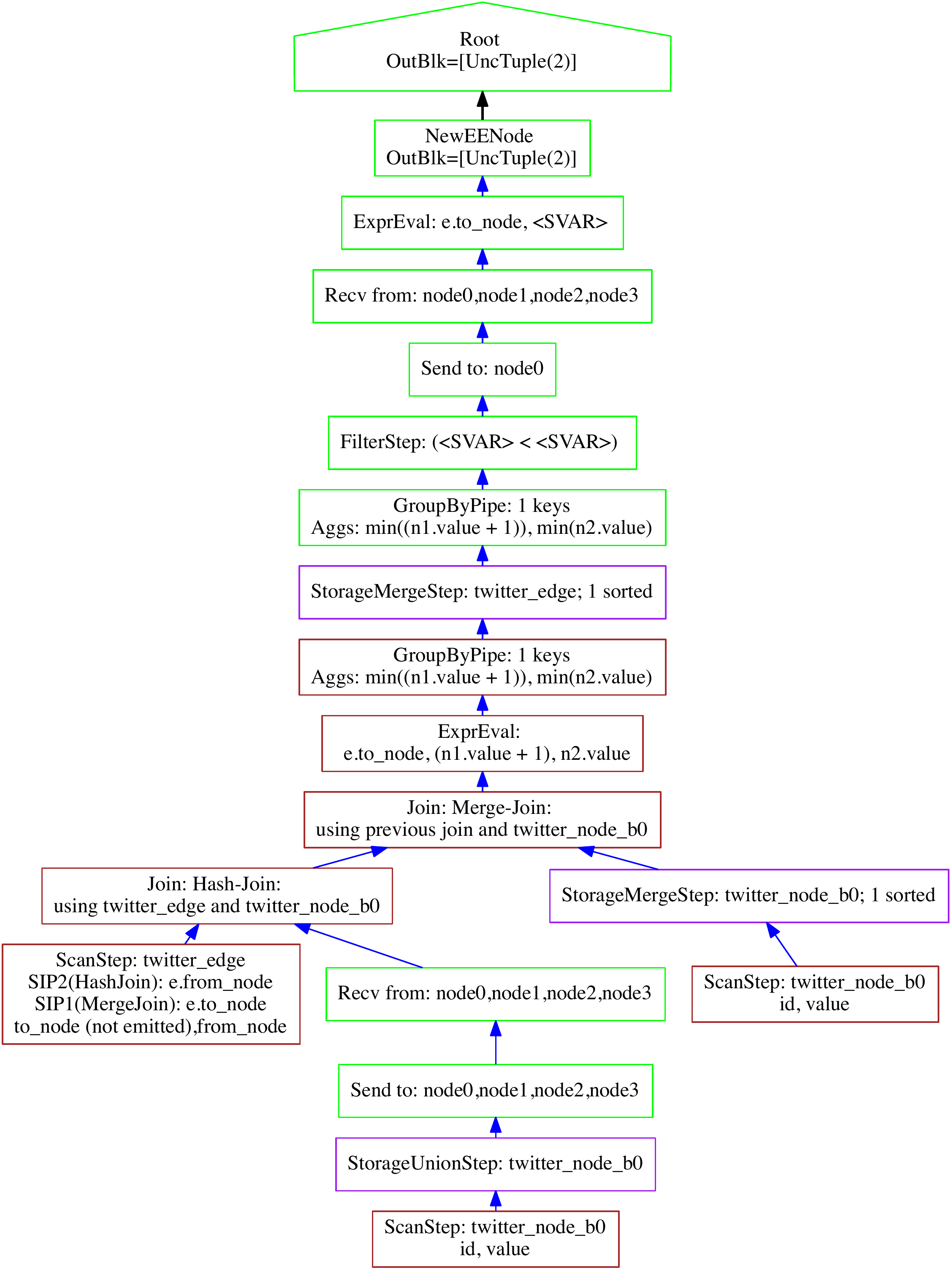}
\vspace{-0.5cm}
\caption{Query execution plan of shortest paths in Vertica.}
\vspace{-.4cm}
\label{fig:vertica-sssp-plan}
\end{figure}

\subsubsection{Intra-query Parallelism}
\label{ref:opt_Parallel_Graph_Exploration}

Vertica includes capabilities that allow it employ multiple cores to process a single query.  To allow Vertica to explore graphs in parallel as much as possible, we rewrite graph exploration queries that involve a self-join on the edges table by adding a GROUP BY clause on the edge id, and let Vertica partition the groups across CPU cores to process subgraphs in parallel.
Though such parallel graph exploration ends up doing more work in each iteration,  it still reduces the number of joins and results in much better performance.

\subsubsection{Example SSSP Query Execution on Vertica}

We now look at the a specific example of physical query execution plan for single source shortest path (SSSP) on Vertica. Figure~\ref{fig:vertica-sssp-plan} shows the plan for running SSSP over the Twitter graph (41 million nodes, 1.4 billion edges). The query involves two joins, one hash and the other sort merge. To perform the hash join, the system broadcasts the smaller node relation (shown in the middle).
While scanning the large edge relation, the query applies two SIP filters, one for the hash join condition and other for the merge join condition, in order to filter unnecessary tuples at the scan step itself.
The hash join blocks on the smaller node relation. However, once the hash table is built, the remainder of the query is fully pipelined, including the merge join, the group by, and writing the final output. This is possible because the merge join and the group by are on the same key. As a result, data does not need to be re-segmented for the group by and the system performs 1-pass aggregation locally on each machine.

The above query execution plan is different from the Giraph query execution pipeline of Figure~\ref{fig:giraphplan} in three ways: (i)~it filters the unnecessary tuples from the large edge table as early as possible by using sideways information passing, (ii)~it fully pipelines the query execution as opposed to blocking the data flow at the vertex function in Giraph, and (iii)~it picks the best join execution strategies and broadcasts the data wherever required as compared to the static hard-coded join implementation in Giraph. As a result, Vertica is able to produce better execution strategies for such graph queries.


\subsection{Extending Vertica}
\label{sec:udfs}
Relational database are extensible by design via the use of UDFs.
We see how we can extend Vertica to address two issues: (i)~how to run unmodified vertex programs without translating to SQL, and (ii)~avoiding the expensive intermediate disk I/O in iterative graph queries.

\subsubsection{Running Unmodified Vertex Programs}

\begin{figure}[!t]
\hspace{-0.3cm}
\includegraphics[width=3.5in]{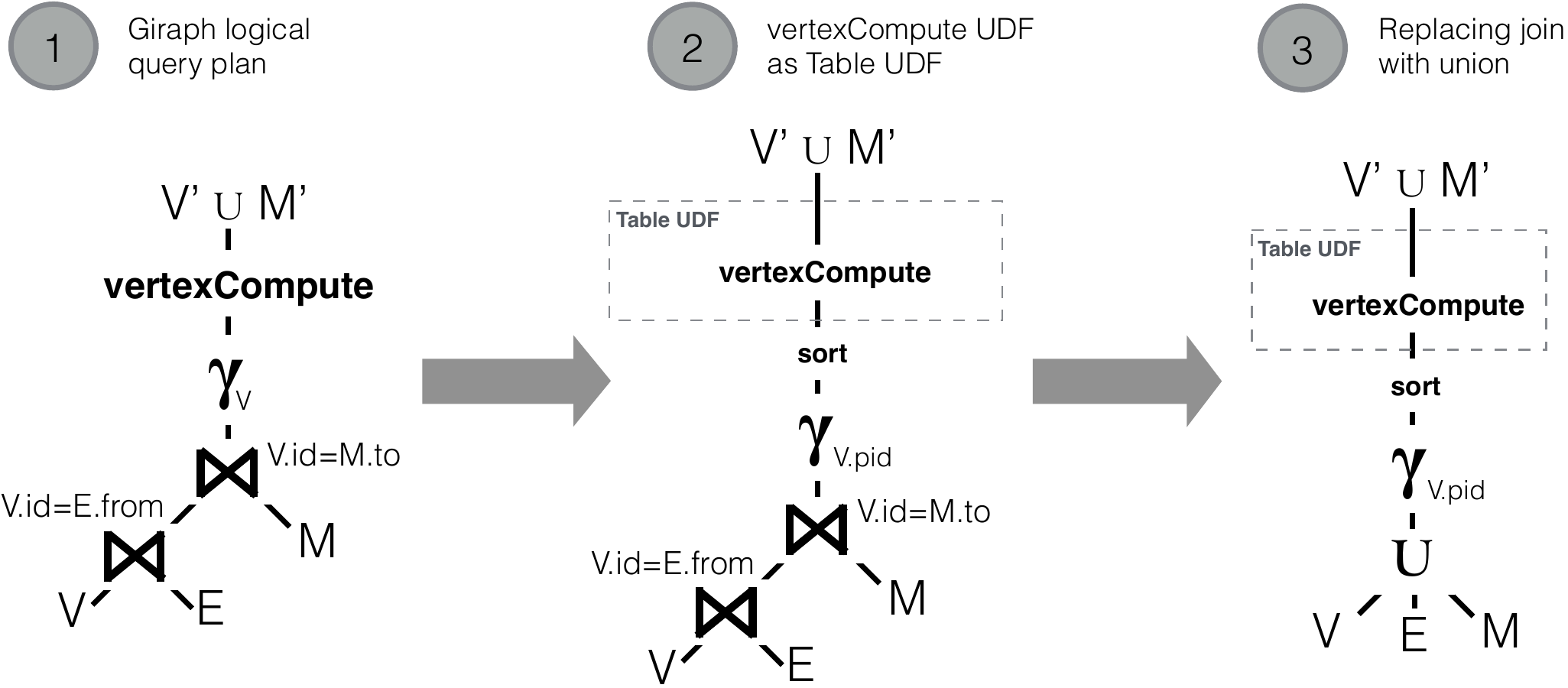}
\vspace{-0.6cm}
\caption{Rewriting Giraph Query Plan using Table UDFs.}
\vspace{-0.5cm}
\label{fig:giraphquery_vertexcentric}
\end{figure}

We saw in Section~\ref{section:sqltranslation} that several common vertex functions can be rewritten as relational operators. However, certain algorithms, such as collaborative filtering, have more sophisticated vertex function implementations which cannot easily be mapped to SQL operators. We can, however, still run vertex functions as table UDFs in Vertica without translating to relational operators. The middle of Figure~\ref{fig:giraphquery_vertexcentric} shows such a logical query plan.
We first partition the vertices, then sort the vertices in each partition, and finally invoke the table UDF for each partition.
The table UDF iterates over each vertex, invokes the \textit{vertexCompute} function over it, and outputs the union of updated vertices ($V'$) and outgoing messages ($M'$).
As an optimization, by batching several vertices in each table UDF, we can significantly reduce the UDF overhead in relational databases.
This query plan can be further improved by replacing the table joins with unions, as shown in the right of Figure~\ref{fig:giraphquery_vertexcentric}. The table UDF is then responsible for segregating the tuples from different tables before calling the vertex function.

\subsubsection{Avoiding Intermediate Disk I/Os}
\label{opt:sharedmemory}

Iterative queries generate a significant amount of intermediate data in each iteration.
Since relational databases run iterative queries via an external driver program, the output of each iteration is spilled to disk, thereby resulting in substantial additional I/O.
This I/O is also happens when running vertex functions as table UDFs.
However, we can implement a special table UDF in which the UDF instances load the entire graph at the beginning and store the graph in shared memory, without writing the output of each iteration to disk (emulating the Giraph-like map-only behavior using table UDFs in Vertica).  
The right side of Figure~\ref{fig:giraphquery_vertexcentric} shows such a query plan.
Of course this approach has less I/O at the cost of a higher memory footprint.
And since the entire graph analysis runs as a single transaction, many of the database overheads such as locking, logging, and buffer lookups are further reduced.
However, the UDF is now responsible for materializing and updating $V\Join E$, as well as propagating the messages from one iteration to the other (the synchronization barrier).
Still, once implemented\footnote{Our current implementation runs on multiple cores on a single node. Future work will look at distributing it across several nodes.}, the shared memory extension 
allows users to run unmodified vertex programs (or those which are difficult to translate to SQL).  In some cases it can also
yield a significant speed-up (up to $2.6$ times) over even native SQL variants (as shown in our experiments). 


\begin{figure}[!t]
\hspace{-0.2cm}
\includegraphics[width=3.6in]{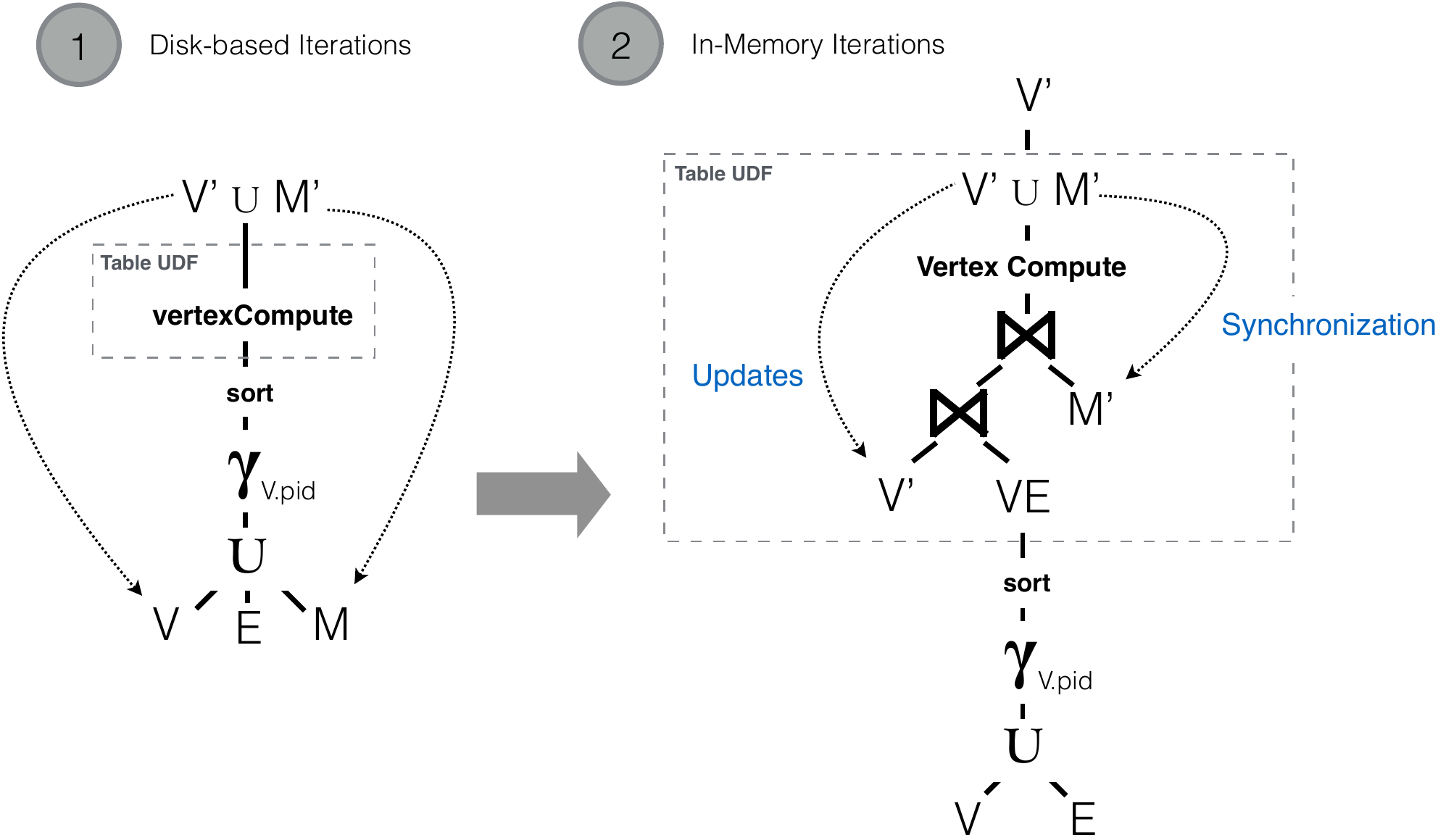}
\vspace{-0.6cm}
\caption{In-memory Vertex-centric Query Execution in Vertica.}
\vspace{-0.5cm}
\label{fig:giraphquery_sharedmemory}
\end{figure}

\begin{figure*}[!t]
\hspace{-0.6cm}
\subfigure[PageRank]{
\includegraphics[height=1.4in]{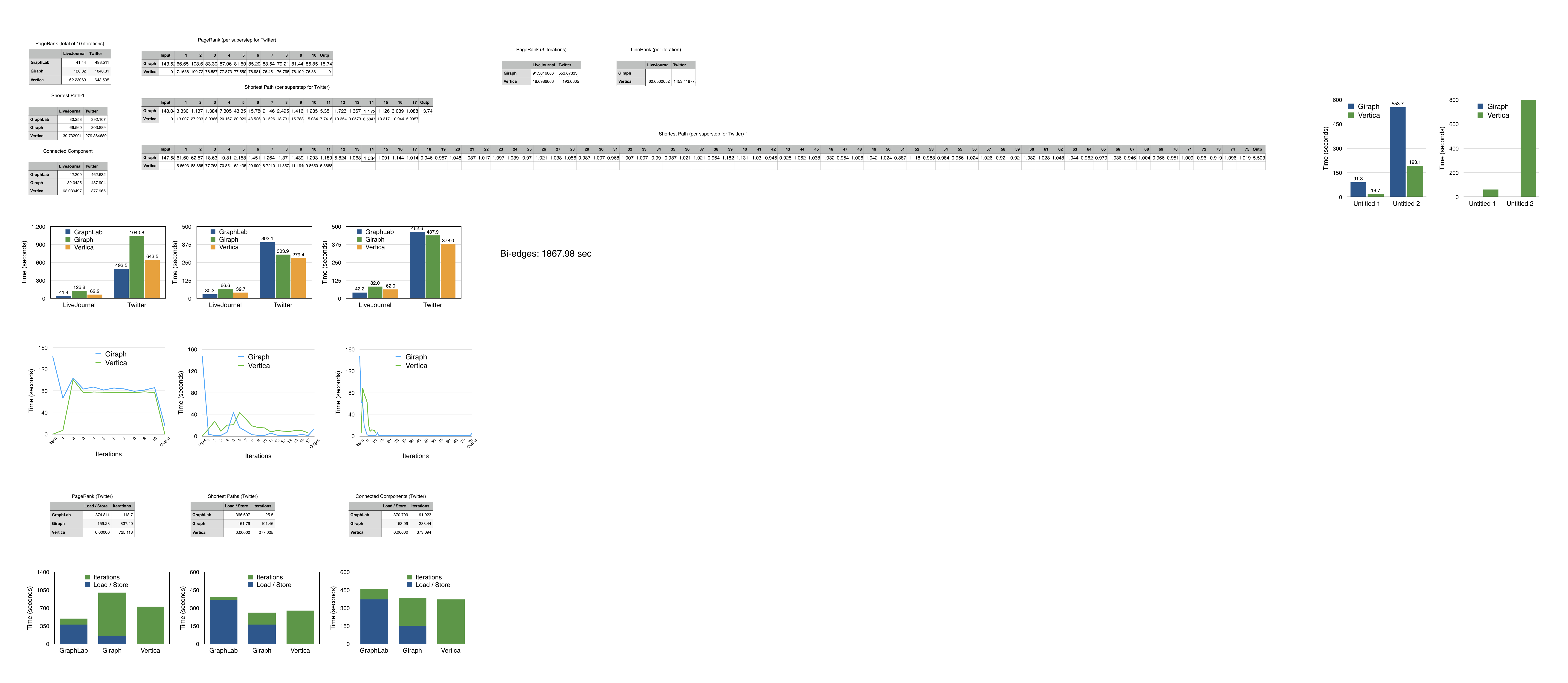}
\label{fig:vertex-pr}
}
\hspace{-0.2cm}
\subfigure[Shortest Path]{
\includegraphics[height=1.4in]{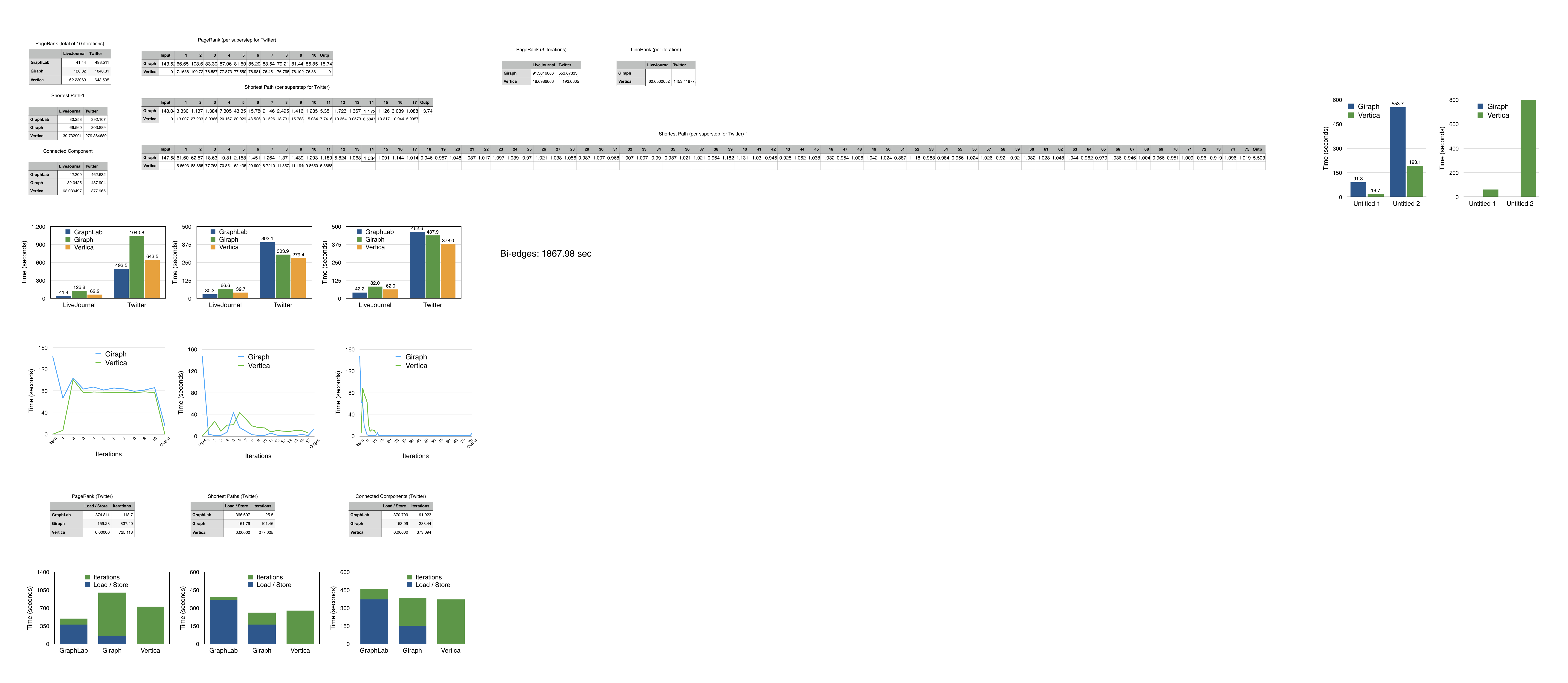}
\label{fig:vertex-sssp}
}
\hspace{-0.2cm}
\subfigure[Connected Components]{
\includegraphics[height=1.4in]{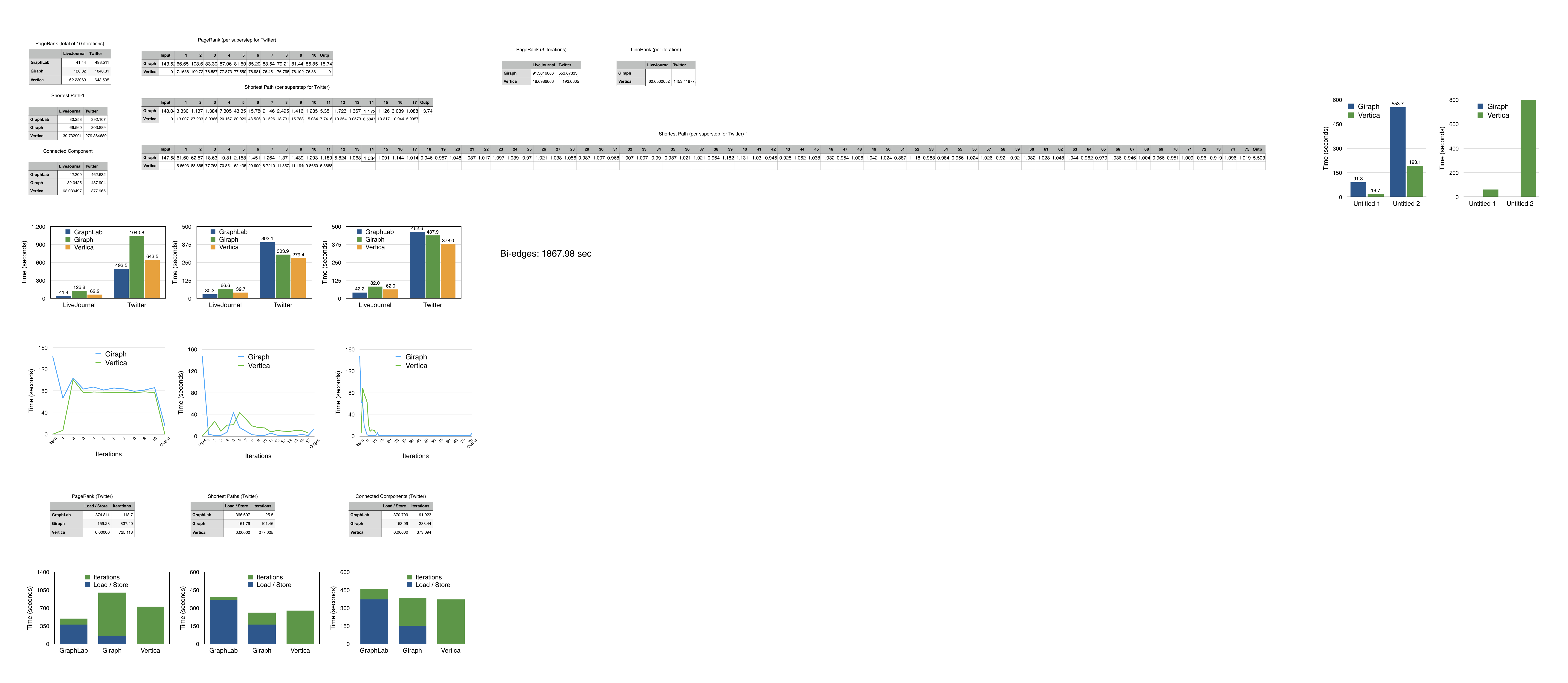}
\label{fig:vertex-cc}
}
\vspace{-0.4cm}
\caption{Typical vertex-centric Analysis Using Vertica.}
\end{figure*}

\begin{figure*}[!t]
\hspace{-0.6cm}
\subfigure[PageRank]{
\includegraphics[height=1.45in]{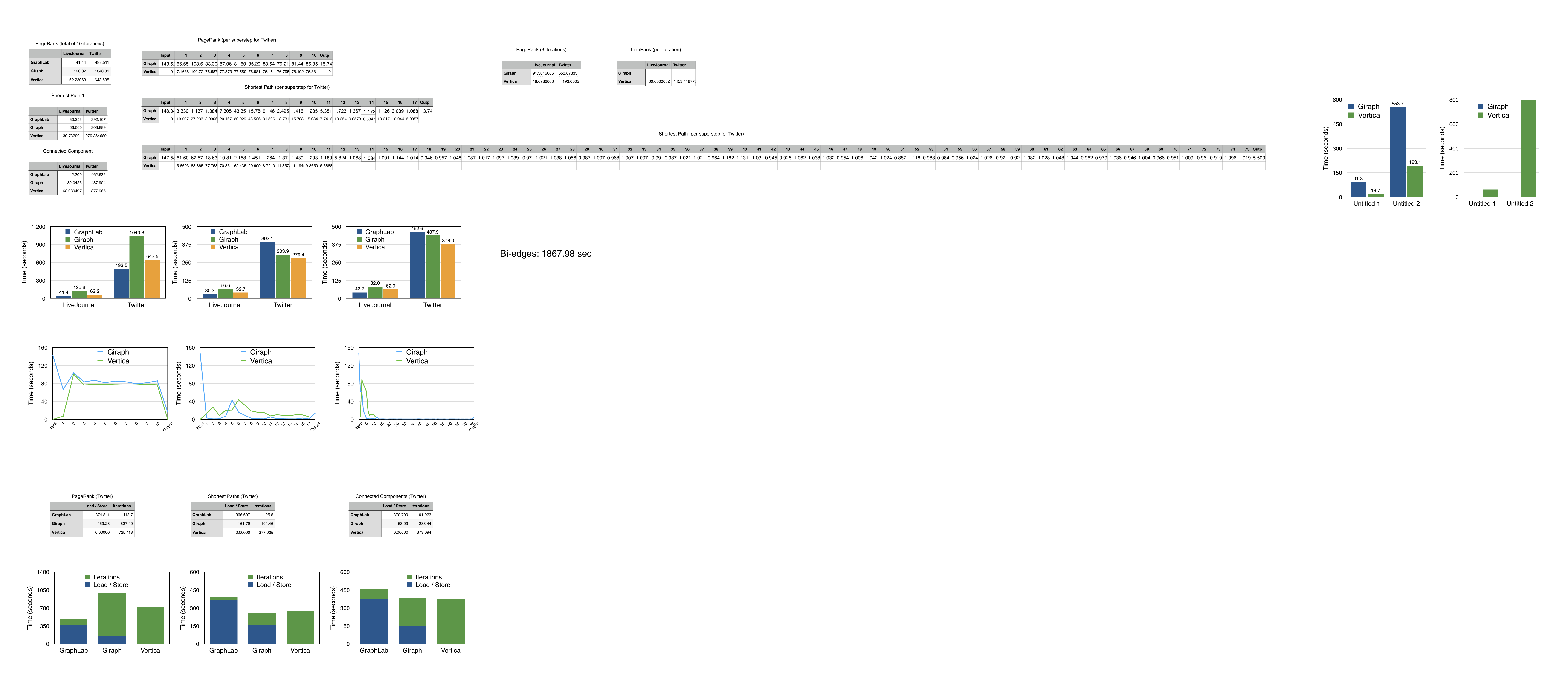}
\label{fig:vertex-pr-itr}
}
\hspace{-0.2cm}
\subfigure[Shortest Path]{
\includegraphics[height=1.45in]{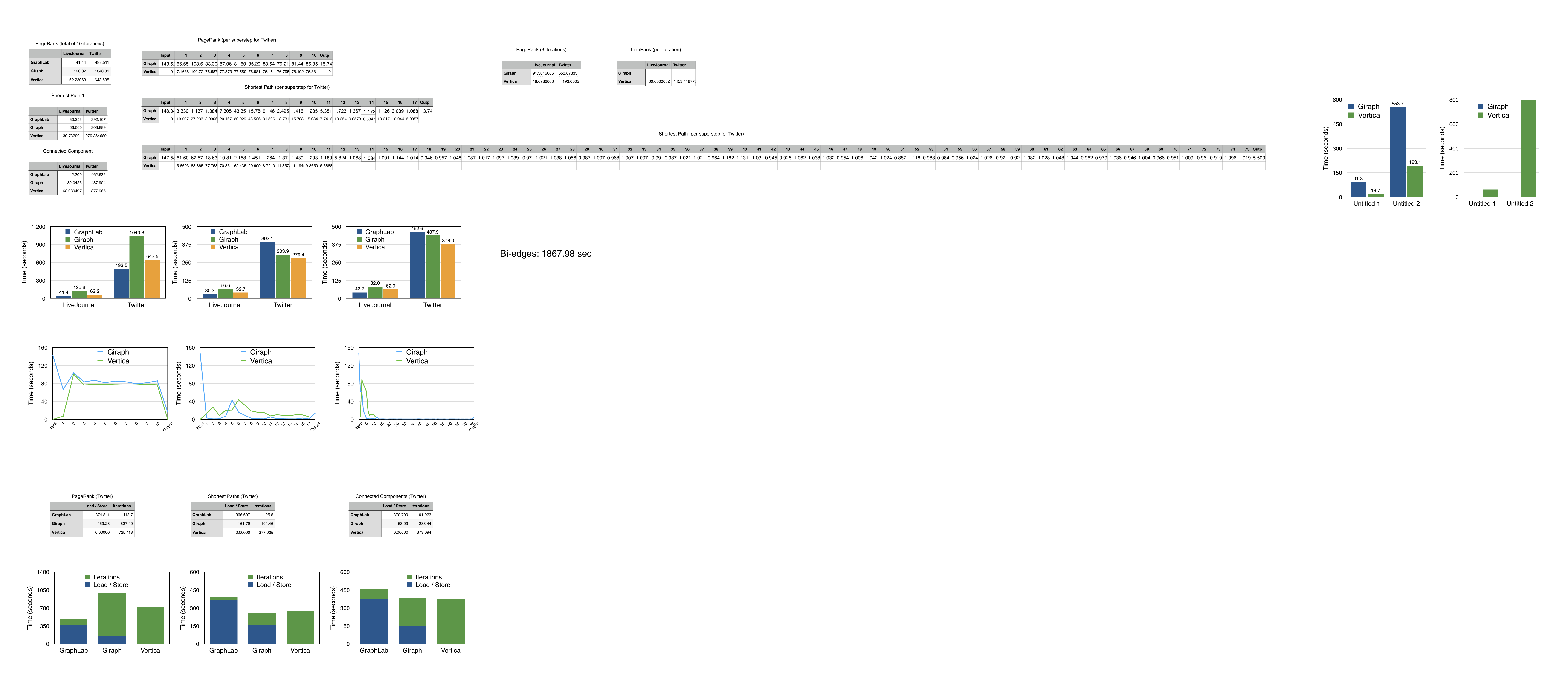}
\label{fig:vertex-sssp-itr}
}
\hspace{-0.2cm}
\subfigure[Connected Components]{
\includegraphics[height=1.45in]{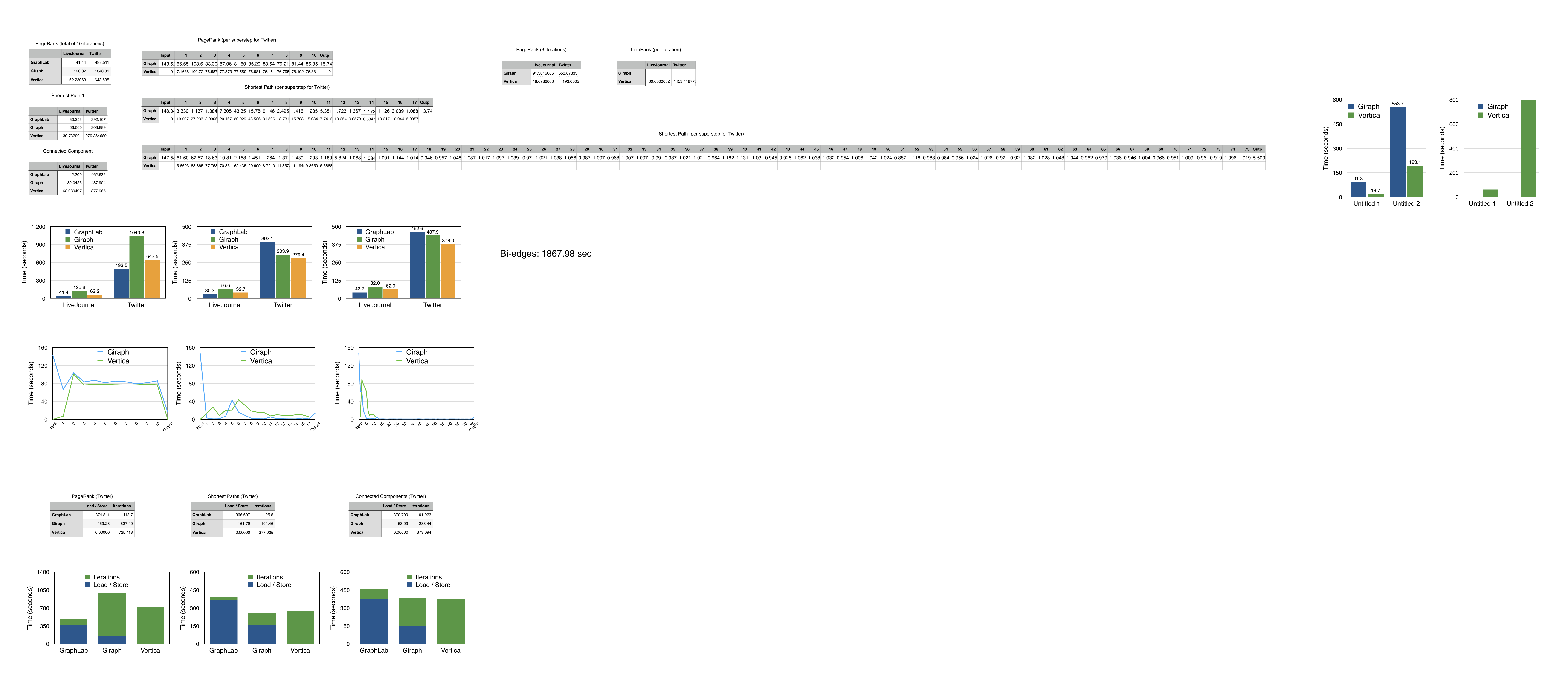}
\label{fig:vertex-cc-itr}
}
\vspace{-0.4cm}
\caption{Per-iteration runtime of typical vertex-centric Analysis on Twitter graph.}
\end{figure*}

\begin{figure*}[!t]
\hspace{-0.4cm}
\subfigure[PageRank]{
\includegraphics[height=1.4in]{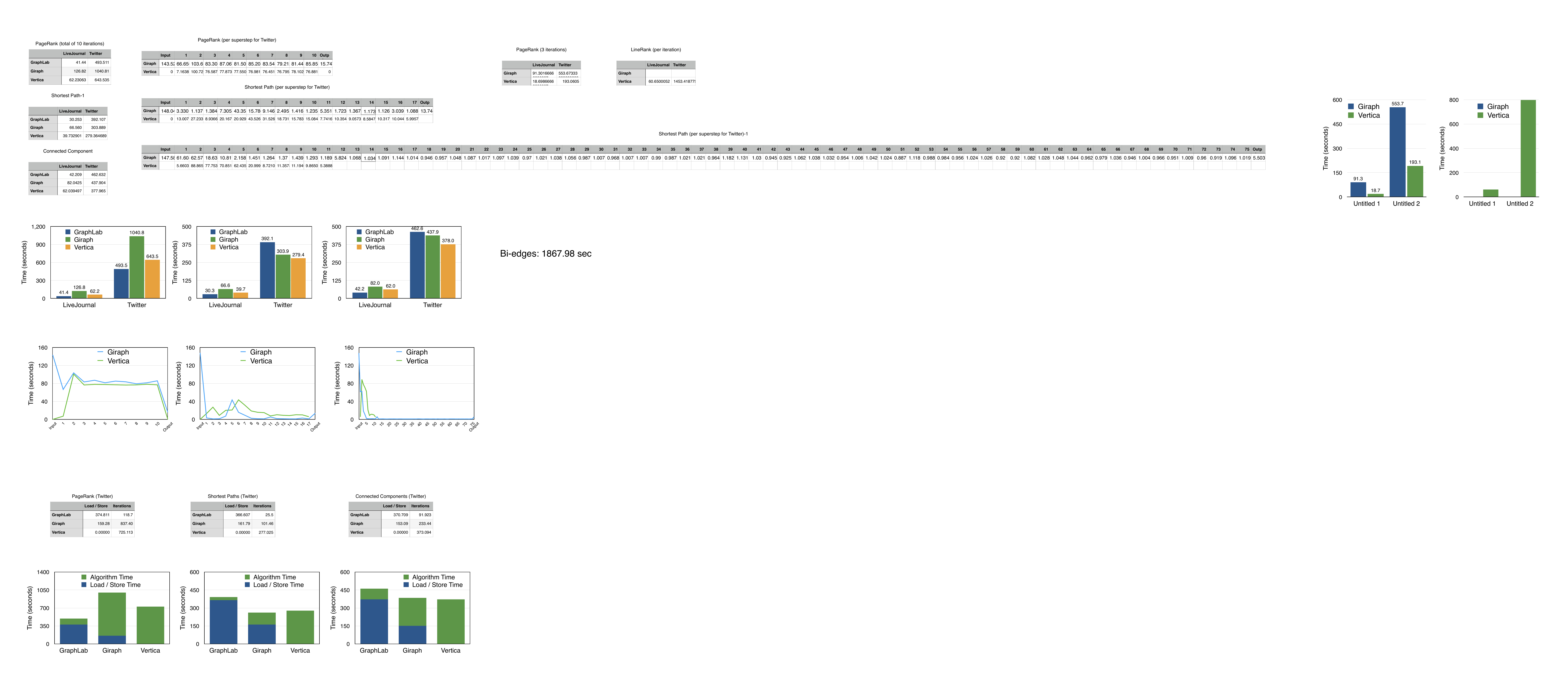}
\label{fig:vertex-pr-breakdown}
}
\hspace{-0.2cm}
\subfigure[Shortest Path]{
\includegraphics[height=1.4in]{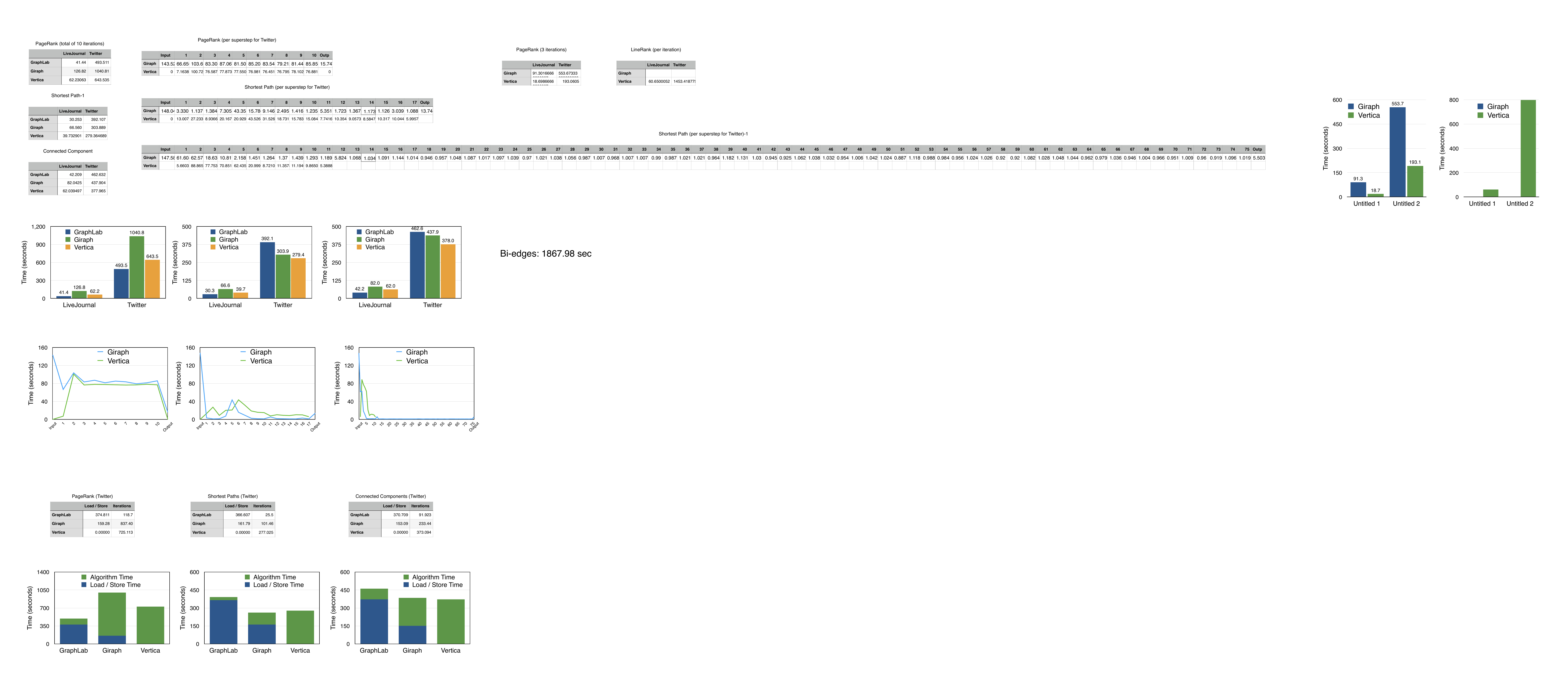}
\label{fig:vertex-sssp-breakdown}
}
\hspace{-0.2cm}
\subfigure[Connected Components]{
\includegraphics[height=1.4in]{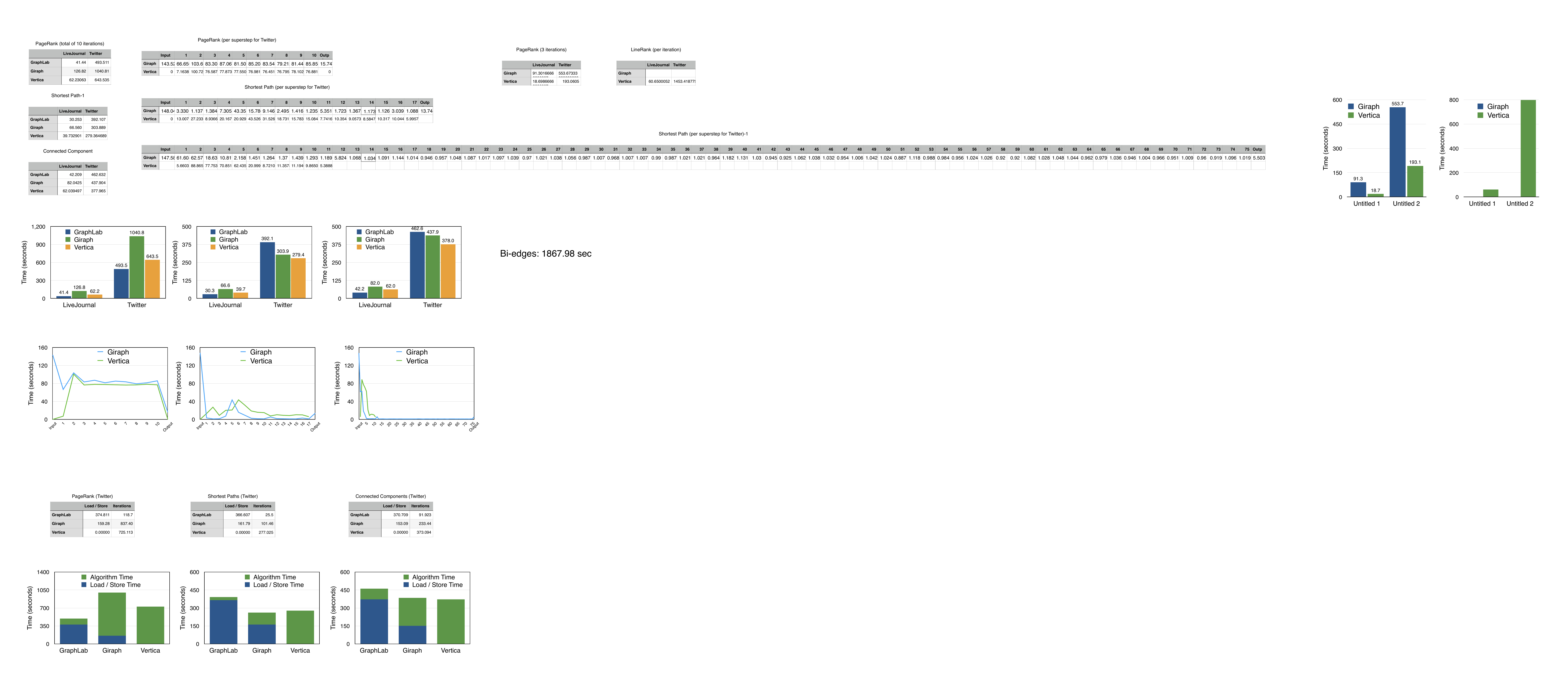}
\label{fig:vertex-cc-breakdown}
}
\vspace{-0.4cm}
\caption{Cost breakdown of typical vertex-centric Analysis on Twitter graph.}
\vspace{-0.2cm}
\end{figure*}

\section{Experiments}
\label{section:benchmarks}


In this section, we describe the experiments we performed to analyze and benchmark the performance of Vertica (version 6.1.2) on graph analytics, and to compare it to dedicated graph-processing systems (Giraph and GraphLab).
We organize our experiments as follows. 
First, we look at the performance of several typical graph queries over large billion-edge graphs. 
Second, we dig deeper and look at the memory footprint and disk I/Os incurred and analyze the differences.
Third, we evaluate our in-memory table UDF implementation of vertex-centric programs over different graph analyses.
Fourth, we study end-to-end graph processing, comprising of  graph algorithms combined with relational operators to sub-select portions of the graph prior to running analytics, project out graph structure from
graph meta-data, and perform aggregations and joins after the graph analysis completes.
Finally, we look at more complex graph analytics beyond vertex-centric analysis, namely 1-hop analysis, and evaluate the utility of Vertica on these operations.

\vspace{0.1cm}
\noindent\textbf{Hardware.} Our test bed consists of a cluster of 4 machines, each having 12 (6x2) 2GHz Xeon cores, running 24-threads with hyper-threading, 48GB of memory, 1.4T disk, running on RHEL Santiago 6.4. We ran all experiments with cold cache and report the average of three runs.

\vspace{0.1cm}
\noindent\textbf{Baselines.} We compare Vertica with two popular vertex-centric graph processing engines, Giraph (version 1.0.0 running on Hadoop 1.0.4 with 4 workers per-node) and GraphLab (version 2.2 running on 4 nodes via MPI and using all available threads).

\vspace{0.1cm}
\noindent\textbf{Datasets.} We ran our benchmarks on a variety of datasets of varying sizes, including both directed as well as undirected graphs. Table~\ref{table:datasets} shows the different datasets used in our evaluation. All datasets are publicly available at \textit{\small http://snap.stanford.edu/data}. 

\vspace{0.1cm}
\noindent\textbf{Data Preparation.} The queries in our experiments must be read data from an underlying data store before running the analysis.
While Vertica reads data from its internal data store, Giraph and GraphLab read the data from HDFS.
All datasets are stored as a list of nodes and a list of edges.
For GraphLab, we further split the data files into 4 parts, such that each node can load (ingress=grid) the graph in parallel during analysis.
Table~\ref{table:datapreparation} summarizes the data preparation costs for the three systems. We can see that Giraph and GraphLab simply copy the raw files to HDFS and load faster  than Vertica. However, Vertica has significantly less disk usage, due to compression and encoding, compared to Giraph and GraphLab.

\vspace{-0.1cm}
\begin{table}[t!]
\center
\begin{scriptsize}
\begin{tabular}{| l | l | r | r |}
\hline
\textbf{Type} & \textbf{Name} & \textbf{Nodes} & \textbf{Edges} \\\hline
\multirow{ 4}{*}{Directed} & Twitter-small & 81,306 & 1,768,149 \\
& GPlus & 107,614 & 13,673,453 \\
& LiveJournal & 4,847,571 & 68,993,773 \\
& Twitter & 41,652,230 & 1,468,365,182 \\\hline
\multirow{ 2}{*}{Undirected} & YouTube & 1,134,890 & 2,987,624 \\
& LiveJournal-undir & 3,997,962 & 34,681,189 \\
\hline
\end{tabular}
\vspace{-0.2cm}
\caption{The datasets used in the evaluation.}
\label{table:datasets}
\end{scriptsize}
\end{table}

\begin{table}[t!]
\center
\begin{scriptsize}
\begin{tabular}{| l | l | r | r | r |}
\hline
\textbf{Metric} & \textbf{Dataset} & \textbf{Vertica} & \textbf{GraphLab} & \textbf{Giraph} \\\hline
\multirow{2}{*}{Upload Time (sec)} & LiveJournal & 45.927 & 15.621 & 12.049\\
& Twitter & 916.421 & 472.358 & 267.799 \\\hline
\multirow{ 2}{*}{Disk Usage (GB)} & LiveJournal & 0.423 & 3.030 & 3.030\\
& Twitter & 9.964 & 73.140 & 73.140 \\
\hline
\end{tabular}
\vspace{-0.2cm}
\caption{Data preparation over two datasets.}
\vspace{-0.4cm}
\label{table:datapreparation}
\end{scriptsize}
\end{table}



\subsection{Typical Vertex-centric Analysis}
\label{subsection:iterativeanalysis}

We first look at the performance of three typical graph queries, namely PageRank, SSSP, and connected components, on Vertica and compare it with Giraph and GraphLab. 
Then, we break the total query time into the time to load/store from disk and the actual graph analysis time.
We used the built-in PageRank, SSSP, and connected component algorithms for Giraph (provided as example algorithms) and GraphLab (provided in the graph analytics toolkit).
For Vertica, we implemented these three algorithms as described below:

\vspace{0.05cm}
\noindent\textbf{PageRank.} We implemented PageRank query shown in Figure~\ref{fig:giraph-querypr_joinelimination} as a combination of two SQL statements on the Vertex (V) and Edge (E) tables: (i)~to compute the outbound PageRank contributed by every vertex:

\vspace{-0.2cm}
\begin{scriptsize}
\begin{alltt}
CREATE TABLE V_outbound AS 
  SELECT  id, value/Count(to_id) AS new_value 
    FROM E,V 
    WHERE E.from_id=V.id 
    GROUP BY id,value;
\end{alltt}
\end{scriptsize}
\vspace{-0.2cm}

\noindent and (ii)~to compute the total PageRank of a vertex as a sum of incoming PageRanks:

\vspace{-0.2cm}
\begin{scriptsize}
\begin{alltt}
CREATE TABLE V_prime AS 
  SELECT to_node AS id, 0.15/N+0.85*SUM(new_value) AS value
    FROM E,V_outbound 
    WHERE E.from_id=V_outbound.id
    GROUP BY to_id;
\end{alltt}
\end{scriptsize}
\vspace{-0.2cm}

\noindent After each iteration, we replace the old vertex table V with V\_prime, i.e.~drop V and rename V\_prime to V. 

\vspace{0.05cm}
\noindent\textbf{Single Source Shortest Path (SSSP).} We implemented SSSP on Vertica using the query shown in Listing~\ref{listing:opt_incremental}, i.e.~we incrementally compute the distances and replace the old vertex table with the new one, unless the number of updates is less than $5000$, in which case we update the vertex table in-place.


\vspace{0.05cm}
\noindent\textbf{Connected Components.} We implemented the HCC algorithm~\cite{pegasus} in Vertica (this is the same algorithm used in Giraph).  HCC  assigns each node to a component identifier. We initialize the vertex values with their ids and in each iteration the vertex updates are computed (similar to SSSP) as follows:

\vspace{-0.2cm}
\begin{scriptsize}
\begin{alltt}
CREATE TABLE v_update_prime AS
  SELECT v1.id, MIN(v2.value) AS value
    FROM v_update AS v2, edge AS e, vertex AS v1
    WHERE v2.id=e.from_node AND v1.id=e.to_node
    GROUP BY v1.id, v1.value
    HAVING MIN(v2.value) < v1.value;
\end{alltt}
\end{scriptsize}
\vspace{-0.2cm}

\noindent As in shortest paths, we apply these updates either in-place or by replacing the vertex table, depending upon the number of updates. 
Additionally, we apply two optimizations: (1)~we do not perform incremental computation at first; rather, we  update all vertices in the first few iterations, i.e.~we use the entire vertex table instead of v\_update in the above query, (2)~since the component ids can be propagated in either edge direction, we propagate them in opposite directions over alternate iterations, i.e.~we change the join condition in the above query to: \texttt{\scriptsize v2.id=e.to\_node AND v1.id=e.from\_node}.  
These optimizations help to significantly speed up the convergence of the algorithm, since component ids can be propagated quickly on either edge directions.


\begin{figure*}[!t]
\hspace{-0.3cm}
\subfigure[GraphLab]{
\includegraphics[height=1.3in]{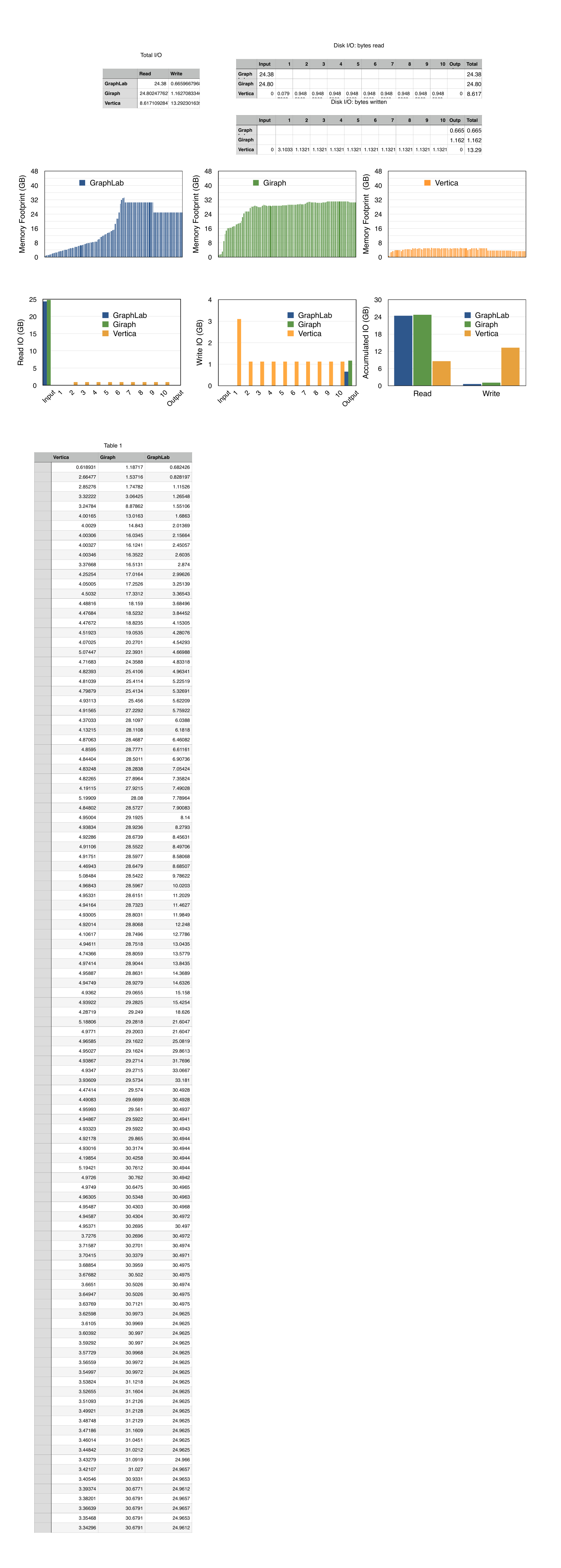}
\label{fig:analysis-mem-graphlab}
}
\hspace{-0.1cm}
\subfigure[Giraph]{
\includegraphics[height=1.3in]{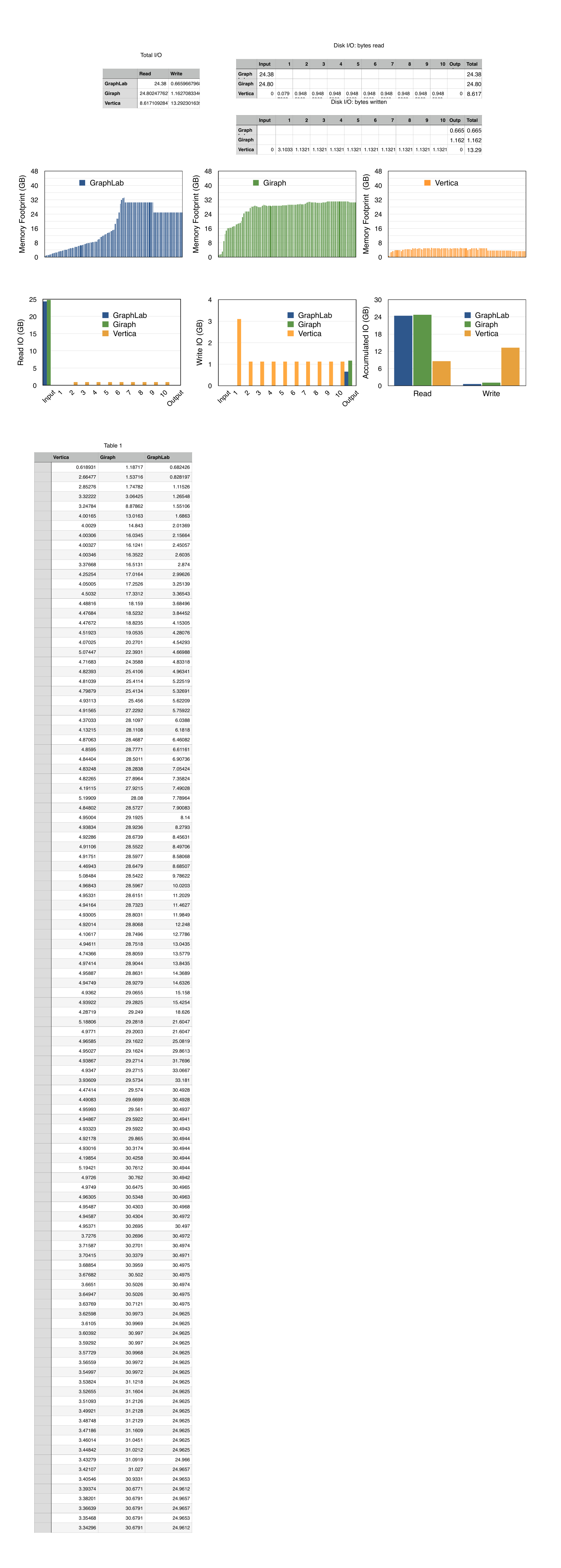}
\label{fig:analysis-mem-giraph}
}
\hspace{-0.1cm}
\subfigure[Vertica]{
\includegraphics[height=1.3in]{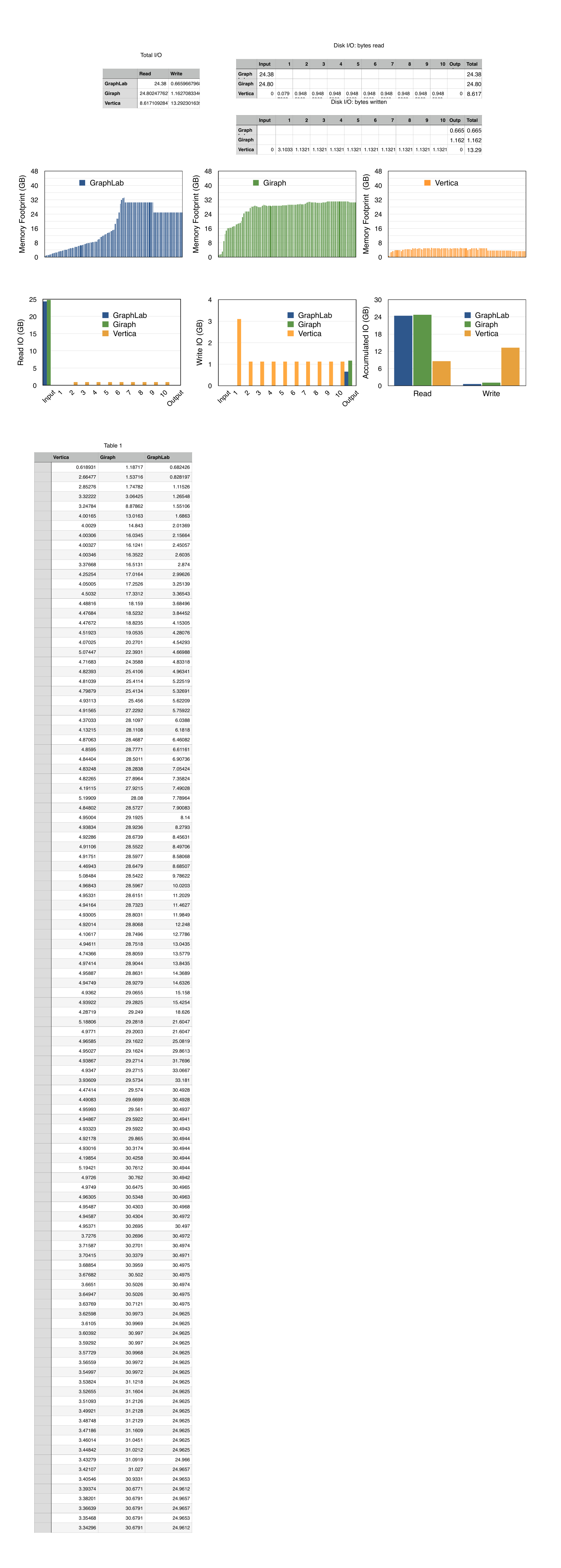}
\label{fig:analysis-mem-vertica}
}
\vspace{-0.4cm}
\caption{The Memory Footprint of Different Systems.}
\end{figure*}

\begin{figure*}[!t]
\hspace{-0.3cm}
\subfigure[Input IO]{
\includegraphics[height=1.55in]{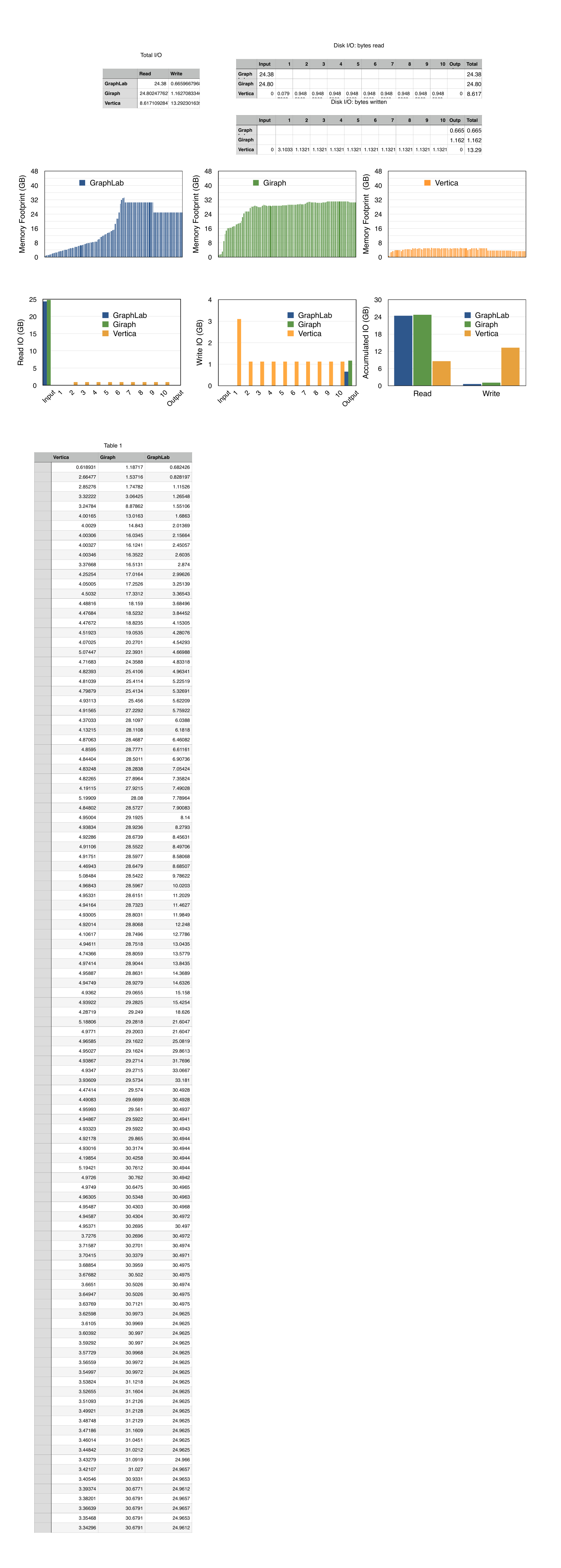}
\label{fig:analysis-io-input}
}
\hspace{-0.2cm}
\subfigure[Output IO]{
\includegraphics[height=1.55in]{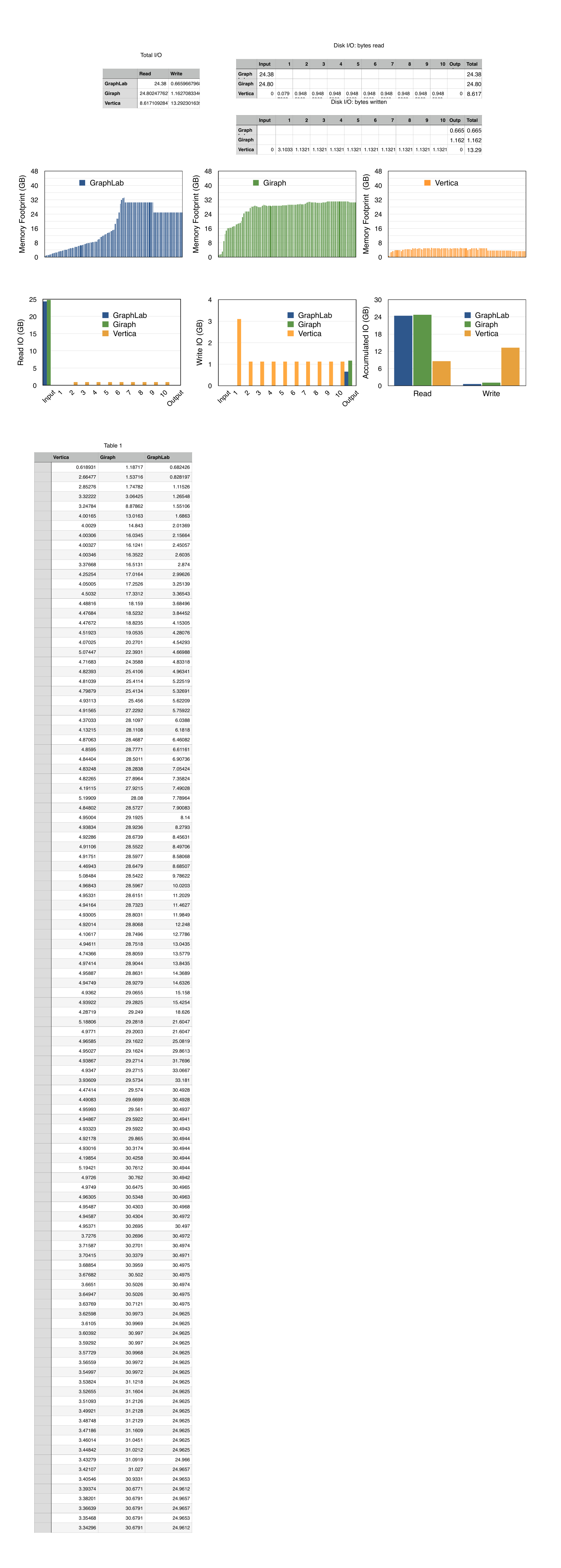}
\label{fig:analysis-io-output}
}
\hspace{-0.2cm}
\subfigure[Total IO]{
\includegraphics[height=1.55in]{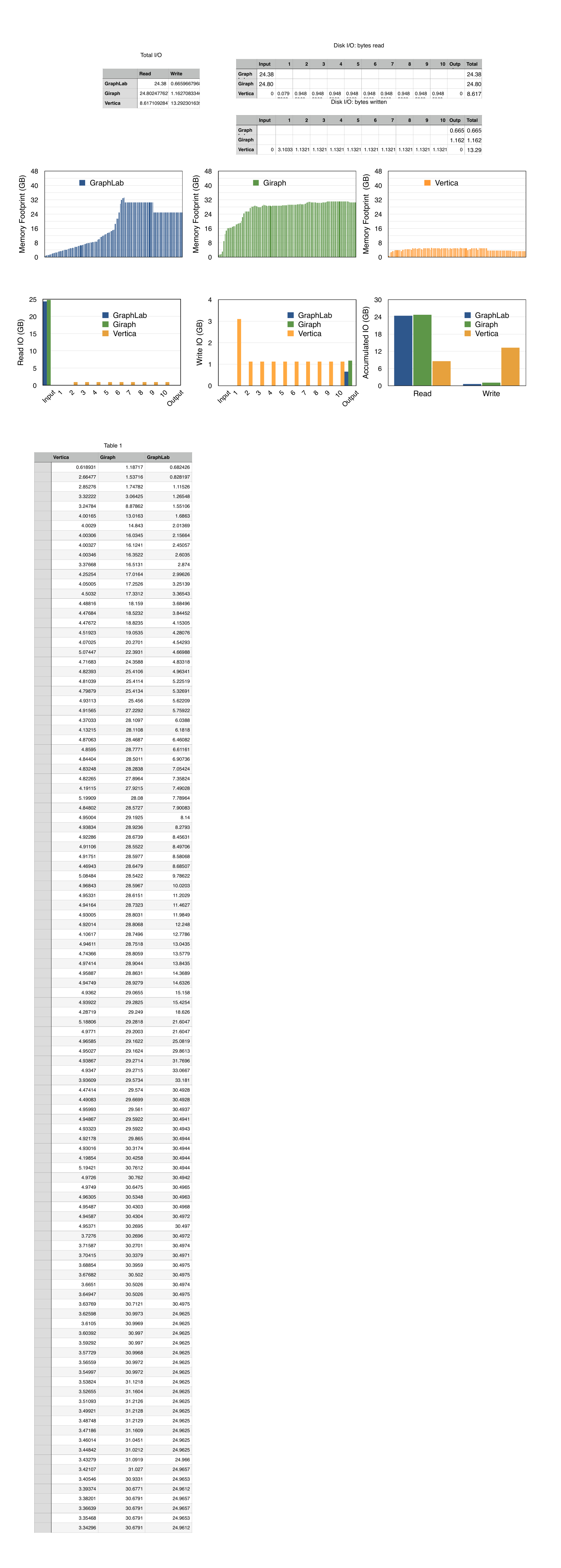}
\label{fig:analysis-io-total}
}
\vspace{-0.4cm}
\caption{The IO Footprint of Different Systems.}
\vspace{-0.2cm}
\end{figure*}

Figures~\ref{fig:vertex-pr},~\ref{fig:vertex-sssp}, and \ref{fig:vertex-cc} show the query runtime of PageRank (10 iterations), SSSP, and Connected Components on GraphLab, Giraph, and Vertica over directed graphs using $4$ machines. We can see that Vertica outperforms Giraph and is comparable to GraphLab on these three queries, both on the smaller LiveJournal graph as well as the billion edge Twitter graph. 
The reason is that these queries are full scan-oriented join queries for which Vertica is heavily optimized.  
A reader might also think that our SQL implementations are thoroughly hand crafted, which is actually true. In fact, one of our goals in this paper is to show that if developers really care about the graph algorithms they can tune them quite a bit in the SQL engines. In Section~\ref{section:inmemorynalysis}, we evaluate our more general shared memory implementation to run arbitrary vertex-centric UDFs.

Note that the advantage of Vertica varies across benchmarks.  For example, on the Twitter graph, Vertica has $40\%$ improvement over Giraph for PageRank, but only $8\%$ improvement for SSSP. To investigate this, consider the per-iteration runtime of these two algorithms, as shown in Figures~\ref{fig:vertex-pr-itr} and~\ref{fig:vertex-sssp-itr}\footnote{The toolkit implementation in GraphLab doesn't produce the per-iteration runtime.}. For PageRank, the iteration runtimes remain relatively unchanged since all vertices are updated in each iteration. For SSSP, the iteration runtime peaks to a maximum since the number of updated vertices first increases and then decreases. From Figures~\ref{fig:vertex-pr-itr} and~\ref{fig:vertex-sssp-itr}, we can see that Giraph incurs very high runtime for the input iteration, which is the cost to load the raw data and transform it into internal graph representation. Vertica has no such input step. Both Vertica and Giraph have similar runtime for iterations having a large number of vertex updates, e.g.~PageRank iterations or the peak iteration in SSSP, due to a full scan over the dataset. However, Giraph has better runtimes for iterations with fewer vertex updates, e.g.~the non-peak iterations of SSSP, due to its in-memory graph representation. Figure~\ref{fig:vertex-cc-itr} shows the per-iteration runtimes of Vertica and Giraph for Connected Components. We can see that by propagating component ids in either edge direction, Vertica converges significantly faster.

Finally, Figures~\ref{fig:vertex-pr-breakdown}, ~\ref{fig:vertex-sssp-breakdown}, and~\ref{fig:vertex-cc-breakdown} show the runtime breakdown of the GraphLab, Giraph, and Vertica on the Twitter graph.
We can see that the total runtime of GraphLab is dominated by the load/store costs. In fact, the analysis time for SSSP in GraphLab is just $25$ seconds, a very small fraction of the overall cost. In contrast, the runtimes of Giraph are dominated by the analysis time.
Vertica, on the other hand, pipelines the data from disk and therefore there is no distinct load/store phase.
However, efficient data storage and retrieval makes Vertica competitive in terms of the total runtime.


\hide{
Lastly, Figure~\ref{fig:vertex-udf-pr} and~\ref{fig:vertex-udf-sssp} show the performance of PageRank (3 iterations) and SSSP, when running as table UDFs in Vertica over different datasets using a single machine.

\vspace{-0.4cm}
\begin{figure}[!h]
\centering
\subfigure[1-node]{
\includegraphics[width=1.5in]{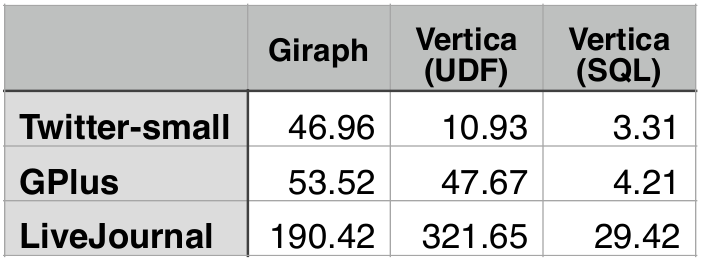}
\label{fig:vertex-udf-pr}
}
\subfigure[4-node]{
\includegraphics[width=1.5in]{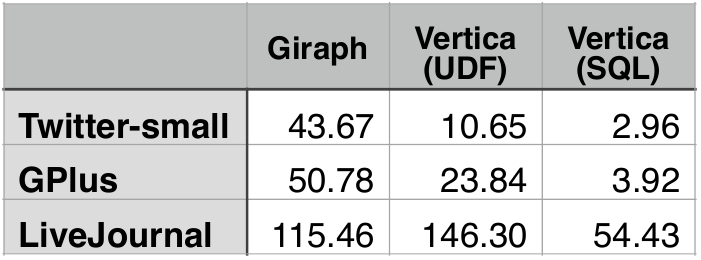}
\label{fig:vertex-udf-sssp}
}
\vspace{-0.2cm}
\caption{Vertex queries as table UDFs in Vertica.}
\vspace{-0.3cm}
\end{figure}

We can see that while the table UDFs are significantly faster than Giraph on smaller graphs, they are a bit slower for larger graphs. This is because in contrast to scan and group by in SQL implementation, the table UDF implementation has too many messages input and output to the UDFs, which typically make the UDFs slow~\cite{trojancolumns}. Thus, the overall runtime of table UDFs is not as good as the SQL implementation. Still, we see that we can run the vertex computations as UDFs and the performance is comparable to Giraph.
}


\subsection{Resource Consumption}

\begin{figure*}[!t]
\hspace{-0.2cm}
\subfigure[Cost Breakdown]{
\includegraphics[height=1.4in]{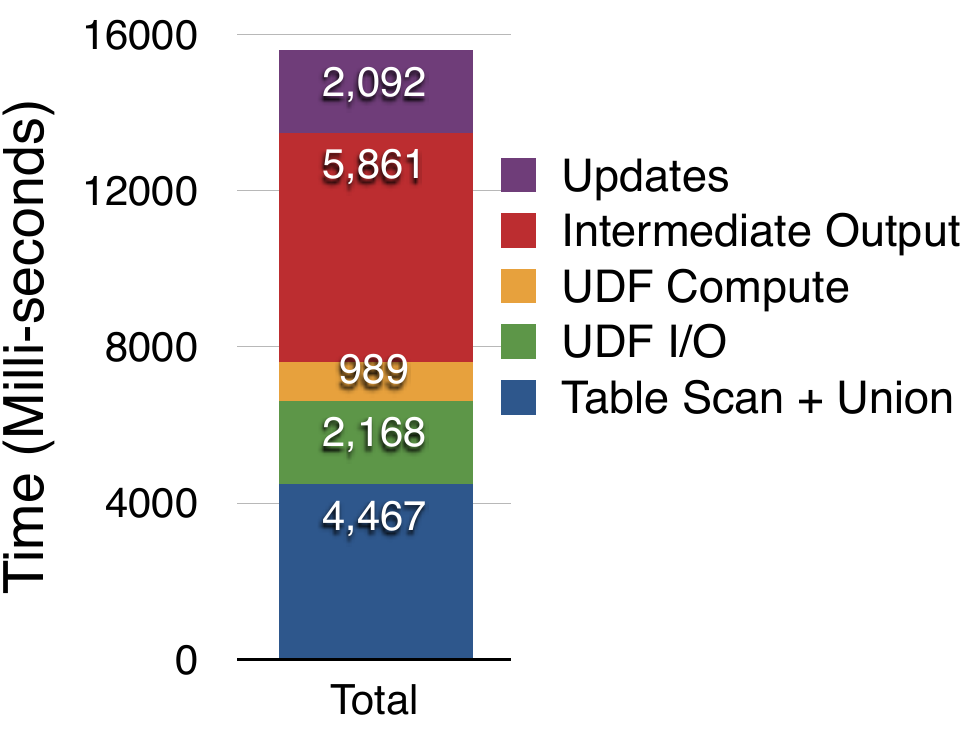}
\label{fig:udf-costbreakdown}
}
\hspace{0.2cm}
\subfigure[Vertex analysis using shared memory]{
\includegraphics[height=1.4in]{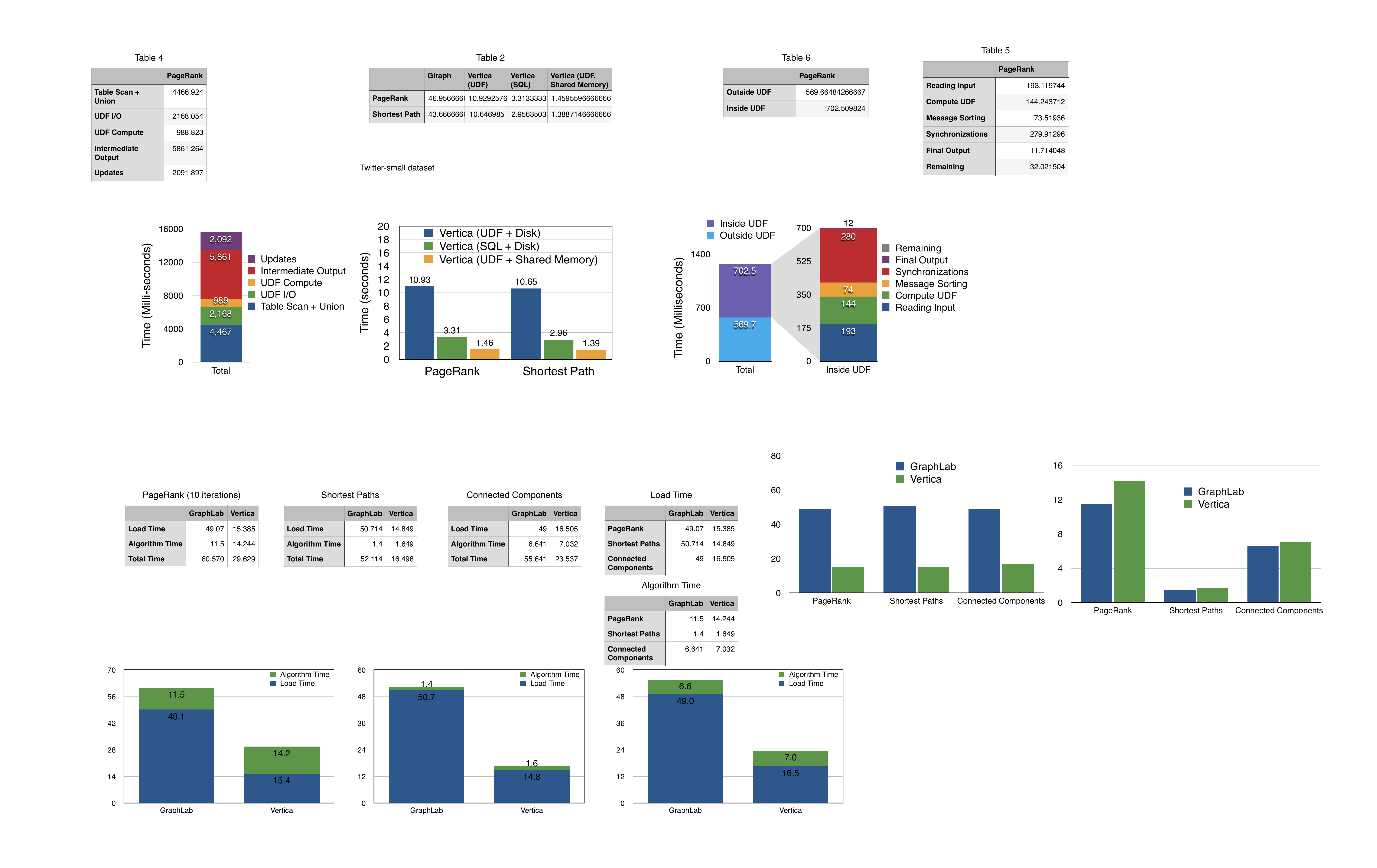}
\label{fig:udf-sharedmemory}
}
\hspace{0.2cm}
\subfigure[Shared memory cost breakdown]{
\includegraphics[height=1.4in]{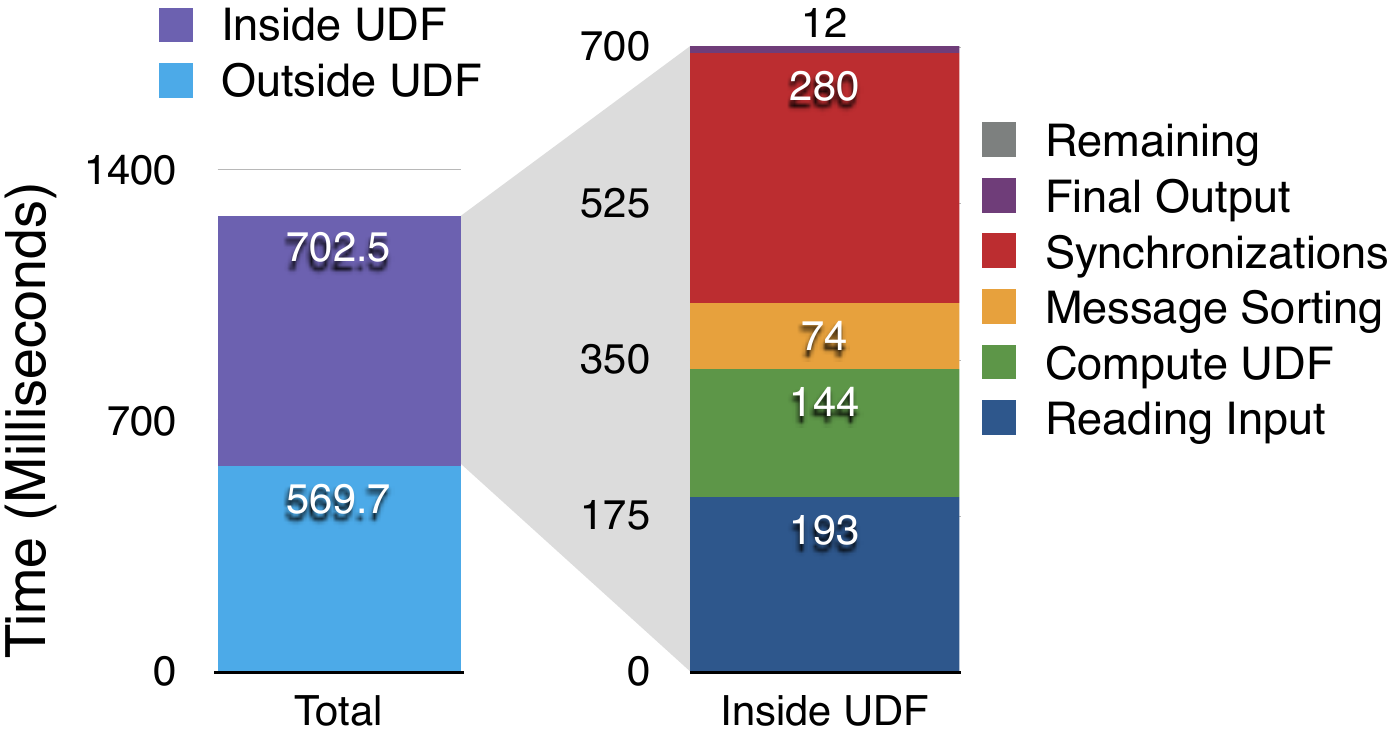}
\label{fig:udf-sm-breakdown}
}
\vspace{-0.4cm}
\caption{Analyzing Resource Consumption and Improving I/O Performance in Vertica.}
\end{figure*}

\begin{figure*}[!t]
\hspace{-0.5cm}
\subfigure[PageRank]{
\includegraphics[height=1.45in]{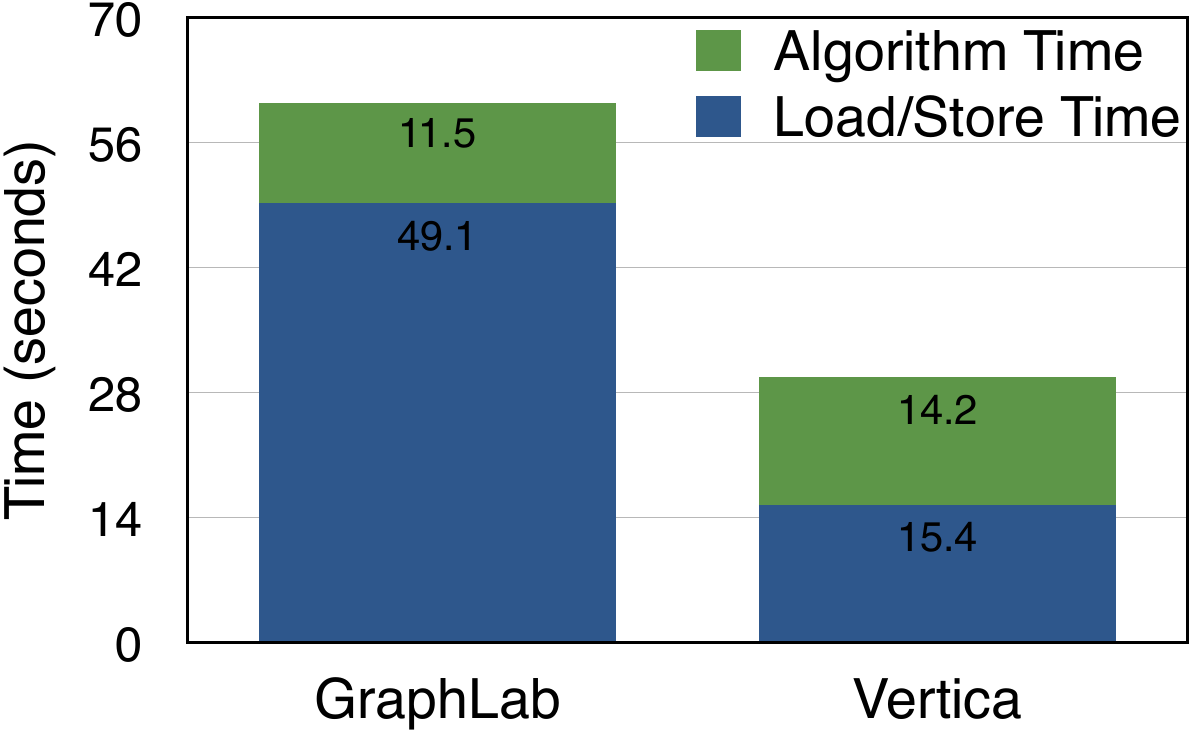}
\label{fig:vertex-sm-pr}
}
\subfigure[Shortest Path]{
\includegraphics[height=1.45in]{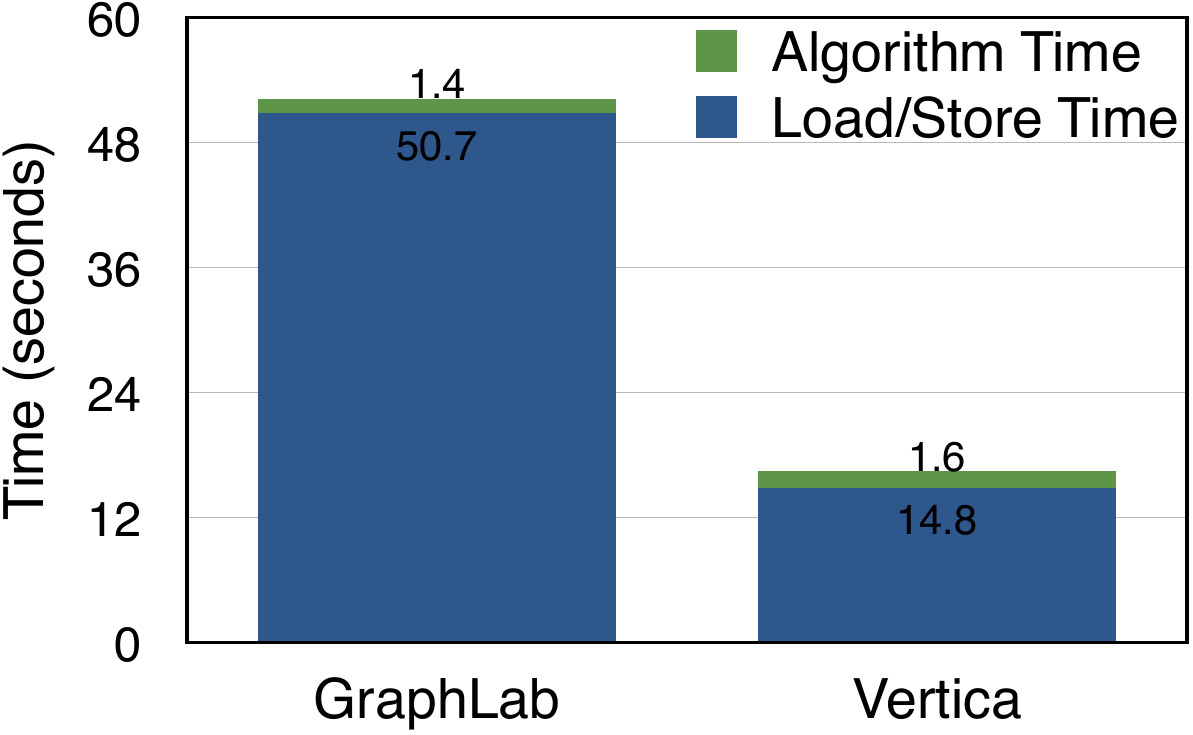}
\label{fig:vertex-sm-sssp}
}
\subfigure[Connected Components]{
\includegraphics[height=1.45in]{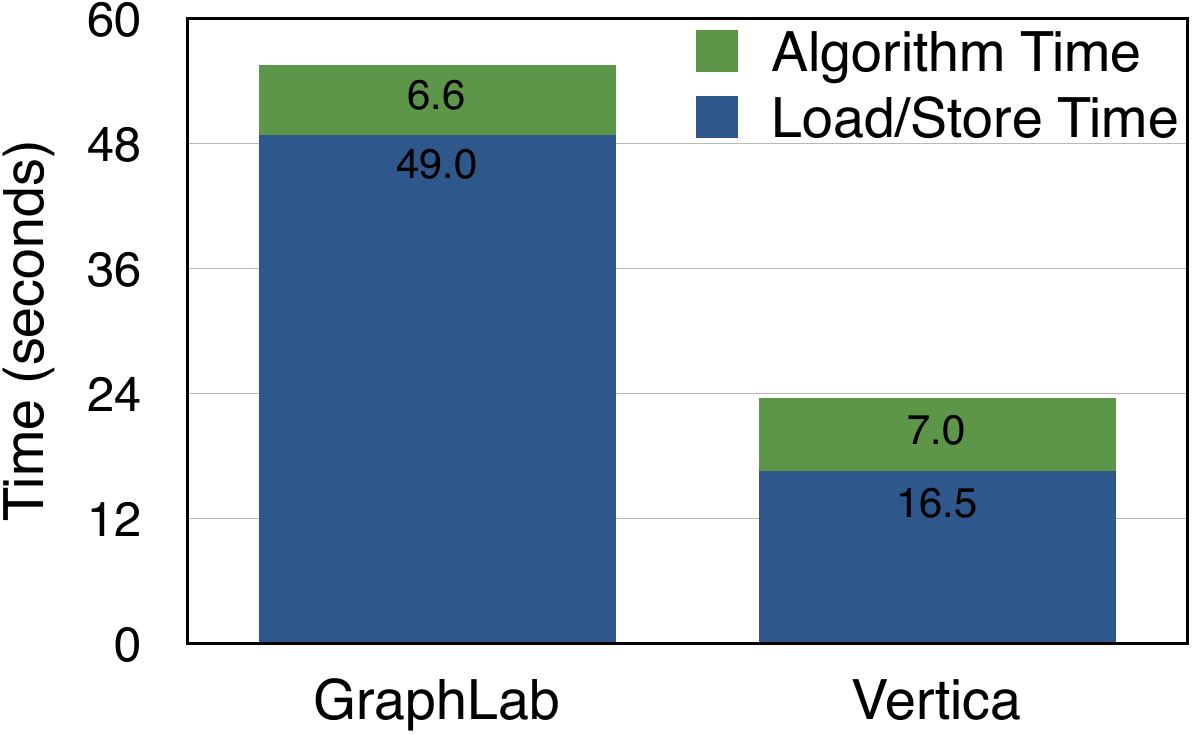}
\label{fig:vertex-sm-cc}
}
\vspace{-0.4cm}
\caption{In-memory Vertex-centric Graph Processing.}
\vspace{-0.2cm}
\end{figure*}

We now study the resource consumption, i.e., the memory footprint and disk I/O, of Vertica, Giraph, and GraphLab. 
Figures~\ref{fig:analysis-mem-graphlab}, ~\ref{fig:analysis-mem-giraph}, and~~\ref{fig:analysis-mem-vertica} show the per-node memory consumption of GraphLab, Giraph, and Vertica when running PageRank on the Twitter graph using $4$ machines.
We can see that out of the total $48$GB memory per node, both GraphLab and Giraph has a peak memory usage of close to $32$GB, i.e., ~$66\%$ of the total memory.
In contrast, Vertica has a peak usage of only $5.2$GB, i.e.~$11\%$ of total memory.
Thus, Vertica has a much smaller memory footprint.

The picture changes completely if we look at the disk I/O. Figures~\ref{fig:analysis-io-input} and~\ref{fig:analysis-io-output} show the number of bytes read and written to disk by GraphLab, Giraph, and Vertica in each PageRank iteration over the Twitter graph.
We can see that GraphLab and Giraph have high read I/O in the input step and no disk reads thereafter, whereas Vertica has no upfront read but incurs disk reads in each iteration. 
Likewise, Vertica incurs high write I/O in each iteration, whereas GraphLab and Giraph incur writes only in the output step.
Thus, in total, while Vertica incurs less read I/O than Giraph (due to better encoding and compression), it incurs much more write I/O (due to materializing the output of each iteration to disk).

This expensive I/O for writing the iteration output is also true when running vertex functions as table UDFs.
To illustrate this, Figure~\ref{fig:udf-costbreakdown} shows the cost breakdown when running PageRank as table UDFs. We can see that writing intermediate output (shown in red) is the major cost in the query runtime.

\subsection{In-memory Graph Analysis}
\label{section:inmemorynalysis}

To test whether we can avoid the expensive intermediate I/O in Vertica, we implemented the shared memory UDFs as described in Section~\ref{opt:sharedmemory}.
Figure~\ref{fig:udf-sharedmemory} shows the runtime of the shared memory table UDF, on the Twitter-small graph.
We can see that the shared memory implementation is about $7.5$ times faster than the disk-based table UDF and even more than $2$ times faster than the SQL implementation.
Finally, Figure~\ref{fig:udf-sm-breakdown} shows the cost breakdown of the shared memory table UDF. We can see that the system spends most of the time inside the UDF and there is no per-iteration output cost.
Thus, we see that developers can indeed sacrifice memory footprint for better I/O performance in Vertica.


We also tested the shared memory Vertica extension with the larger graphs and compared it with GraphLab on a single node.
Figures~\ref{fig:vertex-sm-pr}, ~\ref{fig:vertex-sm-sssp}, and~~\ref{fig:vertex-sm-cc} show the result on the LiveJournal graph.
We can see that the actual graph analysis time (Algorithm Time) of GraphLab and Vertica are very similar now.
However, GraphLab still suffers from the expensive loading time while Vertica benefits from more efficient data storage.
As a result, the performance gap between GraphLab and Vertica widens.
We also scaled SSSP to the billion edge Twitter graph on the shared memory Vertica extension (single node).
The single node algorithm runtime in this case is $44.2$ seconds, which is just $1.7$ times that of GraphLab on 4 nodes\footnote{Single node GraphLab runs out of memory on the Twitter graph.}.
Thus, we see that we can extend Vertica to exhibit similar performance characteristics as main-memory graph engines.

Finally, note that on Vertica, the shared memory runtimes are better than the SQL runtimes (Figures~\ref{fig:vertex-pr},~\ref{fig:vertex-sssp}, and \ref{fig:vertex-cc}) by $2.1$ times on PageRank, $2.4$ times on SSSP, and $2.6$ times on connected components.  This is is largely due to the fact that the shared memory system is able to avoid expensive disk I/O after each iteration.

\subsection{Mixed Graph \& Relational Analyses} Finally, we consider situations when users  want to combine graph analysis with relational analysis.
We extended the graph datasets in our experiment with the following metadata. For each node, we added $24$ uniformly distributed integer attributes with cardinality varying from $2$ to $10^9$, $8$ skewed (zipfian) integer attributes with varying skewness, $18$ floating point attributes with varying value ranges, and $10$ string attributes with varying size and cardinality. For each edge, we added three additional attributes: the weight, the creation timestamp, and an edge type (friend, family, or classmate), chosen uniformly at random. These attributes are meant to model additional relational metadata that would be associated with properties of users in a social media context, or web pages in a search engine. The total size of the Twitter graph with this metadata is $66$ GB. We consider the following end-to-end graph analysis:

\vspace{0.1cm}
\noindent\textbf{(i) Sub-graph Projection \& Selection.} In this analysis, we extract the graph before performing the graph analysis. This includes projecting out just the node ids and discarding all other attributes as well as extracting a subgraph of filtered nodes (e.g., attribute\_6 = 4) connected by edges of type `Family'.
We run PageRank and SSSP over such an extracted subgraph.

\vspace{0.1cm}
\noindent\textbf{(ii) Graph Analysis Aggregation.} In this analysis, we gather the distributions of PageRank (density of nodes by their importance) and distance values (density of neighbors by their distance). This includes computing equi-width histograms after running PageRank and SSSP respectively.


\vspace{0.1cm}
\noindent\textbf{(iii) Graph Joins.} Finally, we combine the output of PageRank and Shortest Paths to emit those nodes which are either very near (path distance less than a given threshold) or are relatively very important (PageRank greater than a given threshold).

\begin{figure*}[!t]
\hspace*{-0.2cm}
\hspace{0.1cm}
\subfigure[Subgraph Projection \& Selection]{
\includegraphics[height=1.4in]{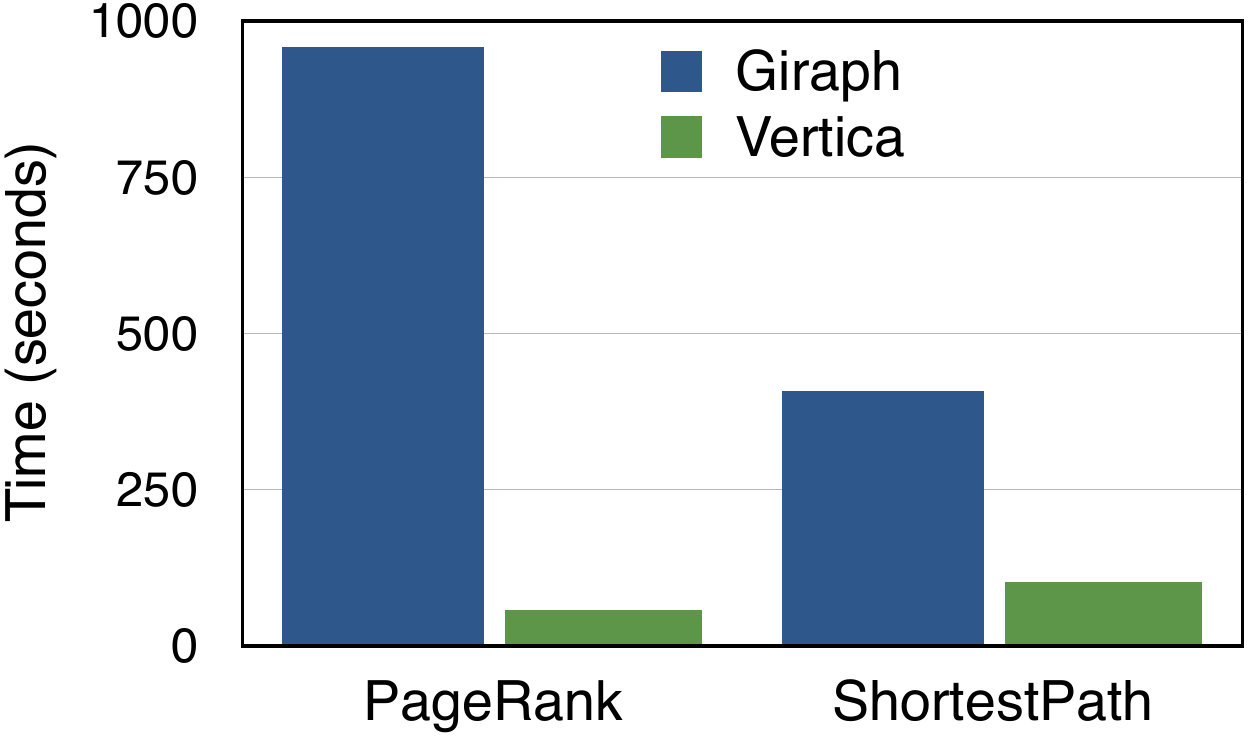}
\label{fig:mixed-selection}
}
\subfigure[Graph Analysis Aggregation]{
\includegraphics[height=1.4in]{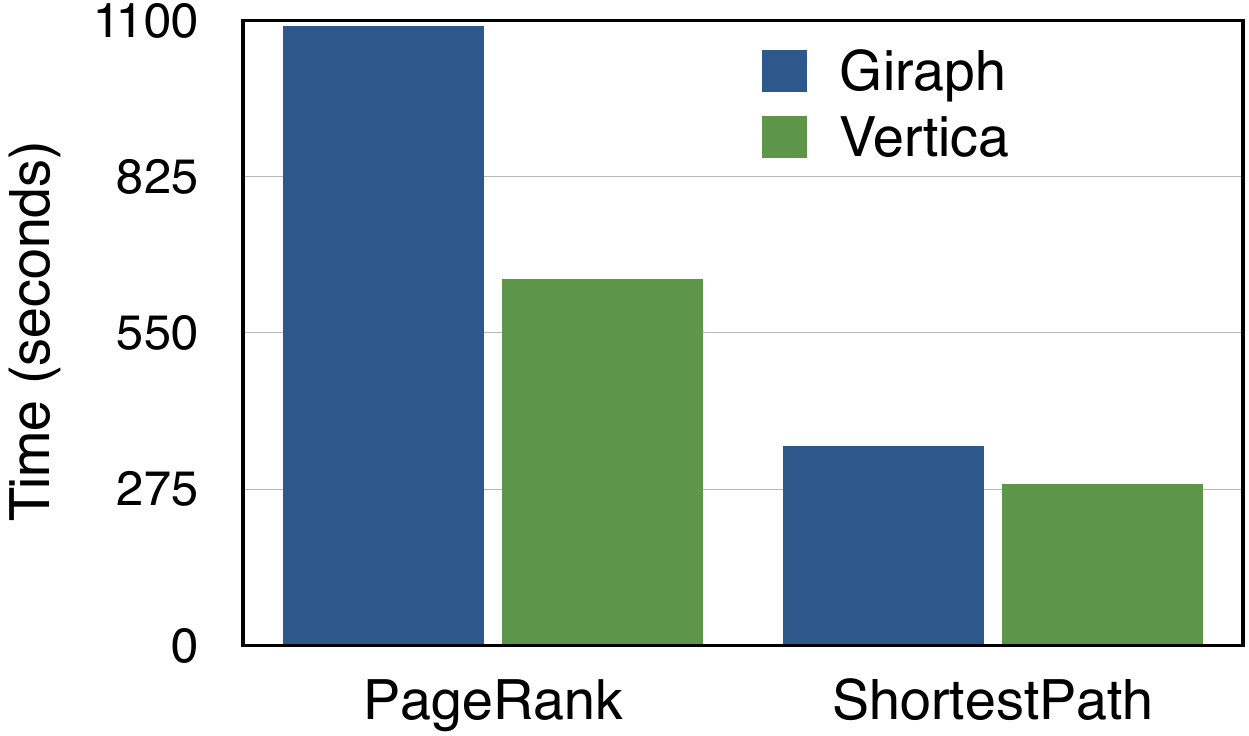}
\label{fig:mixed-aggregation}
}
\subfigure[Graph Join]{
\includegraphics[height=1.4in]{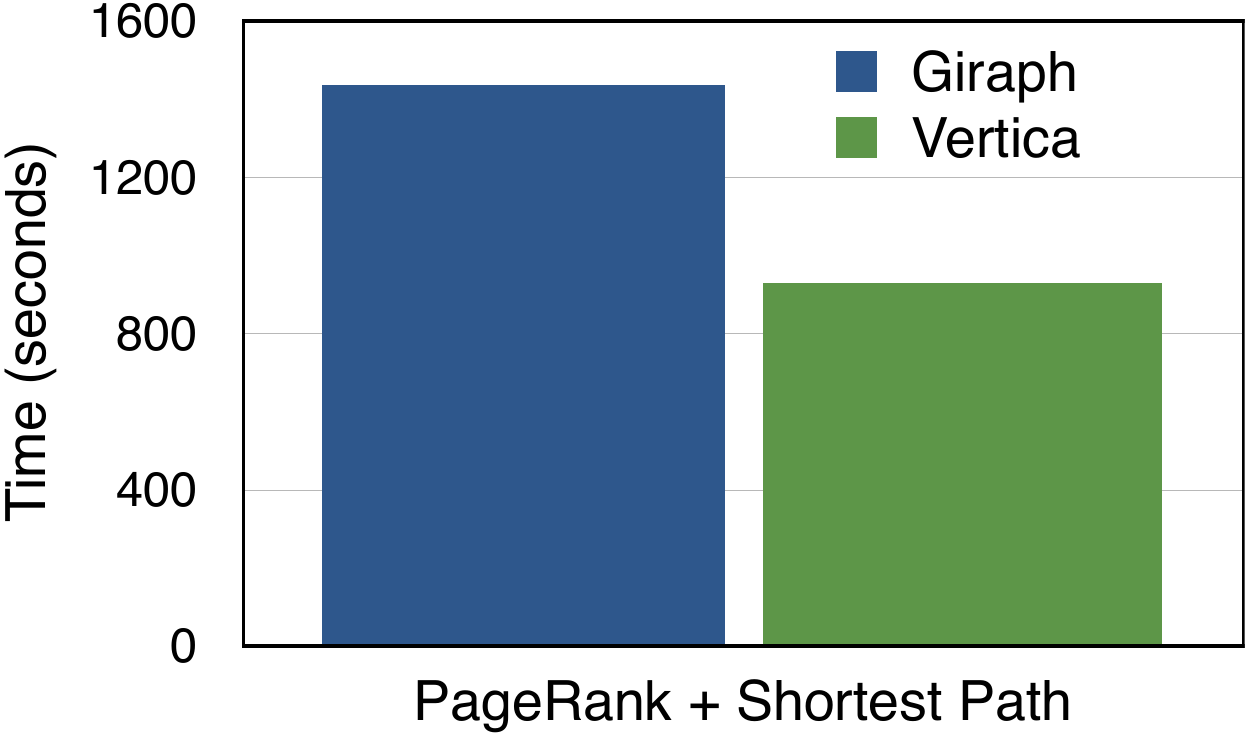}
\label{fig:mixed-join}
}
\vspace{-0.4cm}
\caption{Mixed graph and relational analysis.}
\vspace{-0.2cm}
\end{figure*}

\vspace{0.1cm}
Vertica is well suited for the relational operators in the above three analysis. 
We compare against Giraph, which allows users to provide custom input/output formats that could be used to perform the projection and selection.
We write additional MapReduce jobs for the aggregation and join.
Figures~\ref{fig:mixed-selection} to~\ref{fig:mixed-join} show the result on the Twitter dataset over 4 nodes.
We can see that the performance difference between Vertica and Giraph is much higher now.
For example, Vertica is $17$ times faster on PageRank and $4$ times faster on SSSP, as compared to $1.6$ and $1.08$ earlier, when combining these analysis with sub-graph selection. Performance on Vertica could be further
improved by creating sort orders on the selection attributes.
These massive performance differences are  because Vertica is highly optimized to perform selections and exploit late materialization to access the remaining attributes of only the qualifying records.
In contrast, Giraph incurs a complete sequential scan of data and cannot exploit the fact that part of the analysis is relational in nature.
Rather, developers are required to stitch Giraph programs with programs in another data processing engine such as Hadoop MapReduce or Spark.
Similar results are obtained on aggregation and join queries.

In summary, several data management operations, such as projection, selection, aggregation, and join, fit naturally with graph operations. To perform these operations efficiently, we need several features from relational databases, including efficient layouts, indexes, and statistics, in the graph management system. Column-oriented databases are already well equipped with these techniques, whereas native graphs stores lack optimized implementations of them. As a result Vertica excels on benchmarks that include a mix of relational and graph operations.

\subsection{Beyond Vertex-centric Analyses} 

Let us now look at more advanced graph analytics.
Consider the following two 1-hop neighborhood graph queries.

\vspace{0.1cm}
\noindent\textbf{(i) Strong Overlap.} Find all pairs of nodes having a large number of common neighbors, between them. 
For simplicity, we define overlap as the number of common neighbors. However, we could easily extend the algorithm to include other definitions of overlap. 
Such an analysis to find the strongly overlapping nodes in a graph could be useful in detecting similar entities.
To implement such a query using SQL we do a self-join on the edge table followed by a group by on the two leaf nodes and a count of the number of common neighbors between them. Finally, all those pairs of nodes which have less than the \textit{threshold} number of common neighbors are filtered.

\vspace{-0.2cm}
\begin{scriptsize}
\begin{alltt}
SELECT e1.from_node as n1,e2.from_node as n2, count(*)
 FROM edge e1
 JOIN edge e2 ON e1.to_node=e2.to_node 
  AND e1.from_node<e2.from_node
 GROUP BY e1.from_node,e2.from_node
 HAVING count(*) > THRESHOLD
\end{alltt}
\end{scriptsize}
\vspace{-0.2cm}

\vspace{0.1cm}
\noindent\textbf{(ii) Weak Ties.} Find all nodes that act as a bridge between two otherwise disconnected node-pairs. The goal is to find all such \textit{weak ties} which connect at least a threshold number of node pairs. This is a slightly more complicated query because we need to test for disconnection between node pairs. 
Using SQL, this could be implemented as a three-way join, with the second join being a left join, and counting all cases when the third edge does not exist.

\vspace{-0.2cm}
\begin{scriptsize}
\begin{alltt}
SELECT e1.to_node AS Id, 
 sum(CASE WHEN e3.to_node IS NULL THEN 1 ELSE 0 END)/2 AS C
 FROM edge e1
 JOIN edge e2 ON e1.to_node=e2.from_node 
  AND e1.from_node<>e2.to_node
 LEFT JOIN edge e3 ON e2.to_node=e3.from_node 
  AND e1.from_node=e3.to_node
 GROUP BY e1.to_node
 HAVING C > THRESHOLD
\end{alltt}
\end{scriptsize}
\vspace{-0.2cm}

While the above queries are straightforward to implement and run on Vertica, they are tedious to implement in vertex-centric graph processing systems.
For instance, in Giraph, each vertex needs to broadcast its neighborhood in the first superstep in order to access the 1-hop neighborhood.
Thereafter, we perform the actual analysis in the subsequent supersteps.
Unfortunately, this results in replicating the graph several times in memory.
As an optimization in Giraph, we can reduce the number of messages by sending messages only to those neighbors with a higher node id. 
In addition, we can reduce the size of messages by sending only the ids which are smaller than the id of the receiving vertex.
Still, passing neighborhoods as messages is not natural in the vertex-centric model and incurs the overhead of serializing and deserializing the nodes ids.

\begin{table}[t!]
\center
\begin{scriptsize}
\begin{tabular}{| l | l | r | l |}
\hline
\textbf{Query} & \textbf{Dataset} & \textbf{Vertica} & \textbf{Giraph} \\\hline
\multirow{2}{*}{\it Strong Overlap} & Youtube & 259.56 & 230.01 \\
& LiveJournal-undir & 381.05 & out of memory \\\hline
\multirow{2}{*}{\it Weak Ties} & Youtube & 746.14 & out of memory \\
& LiveJournal-undir & 1,475.99 & out of memory\\
\hline
\end{tabular}
\vspace{-0.2cm}
\caption{1-hop Neighborhood Analysis.}
\vspace{-0.4cm}
\label{table:beyondvertexcentric}
\end{scriptsize}
\end{table}

Table~\ref{table:beyondvertexcentric} shows the performance of Vertica and Giraph over the two 1-hop neighborhood analyses running on $4$ nodes.
We can see that Giraph runs out of memory when scaling to larger graphs (even after allocating $12$GB to each of the $4$ workers on each of the $4$ nodes).
This is because the graphs are quite dense and sending the entire 1-hop neighborhood results in memory usage proportion to the graph size times the average out-degree of a node. Vertica, on the other hand, does not suffer from such issues and works for larger graphs as well. 

In summary, while  Giraph is good for typical graph algorithms such as PageRank and Shortest Paths, it is hard to program and poor in performance for advanced graph analytics such as finding nodes with strong overlap. Vertica, on the other hand is quite facile at such analysis and allows programmers to perform a wide range of analytics, and also yield significantly better performance.

\section{Related Work}

 Existing graph data management systems address two classes of query workloads: (i)~low latency online graph query processing, e.g.~social network transactions, and (ii)~offline graph analytics, e.g.~PageRank computation. Typical examples of systems for online graph processing include RDF stores (such as Jena~\cite{jena} and AllegeroGraph~\cite{allegerograph}) and key value stores (such as Neo4j~\cite{neo4j} and HypergraphDB~\cite{hypergraphdb}). Some recent graph processing systems wrap around relational databases to provide efficient online query processing. These include TAO~\cite{tao} from Facebook and FlockDB~\cite{flockdb} from Twitter, both of which wrap around MySQL to build a distributed and scalable graph processing system. Thus, low latency online graph query processing can be mapped to traditional online processing in relational databases.

Graph analytics, on the other hand, is seen as completely different from traditional data analytics, typically due to its scale and iterative nature. 
As a result a plethora of graph processing systems have been recently proposed.
These include 
vertex-centric systems, e.g.~Pregel~\cite{pregel}, Giraph~\cite{giraph}, GraphLab~\cite{graphlab} and its extensions~\cite{powergraph,graphchi,msync}, GPS~\cite{gps}, Trinity~\cite{trinity}, GRACE~\cite{grace,graceblock}, Pregelix~\cite{pregelix}; 
neighborhood-centric systems, e.g.~Giraph++~\cite{giraph++}, NScale~\cite{nscale,nscale-demo};
datalog-based systems, e.g.~Socialite~\cite{socialite,distsocialite}, GrDB~\cite{grdbdemo,grdb};
SPARQL-based systems, e.g.~G-SPARQL~\cite{gsparql};
and matrix-oriented systems, e.g.~Pegasus~\cite{pegasus}.
Some recent works have also looked at specific graph analysis using relational databases.
Examples include triangle counting~\cite{vertica-triangles}, shortest paths~\cite{relationalsp,oracle-graph}, subgraph pattern matching~\cite{graphiso}, and social network analysis~\cite{vertica-social,benchmark-graph}.
Others have looked at the utility of combining graph analysis with relational operators~\cite{graphx,vertexica}.

In this work, we have a particular emphasis on the question as to whether, {\it in general} relational databases and Vertica in particular can provide a performant, easy-to-use engine on which graph analytics can be layered, as opposed to looking at specific analyses.

\section{Lessons Learned \& Conclusion}

Graph analytics is emerging as an important application area, with several graph data management systems having been recently proposed. These systems, however, require users to switch to yet another data management system. This paper demonstrates that efficient and scalable graph analytics is possible within Vertica relational database system. 
We  implemented a variety of graph analyses on Vertica as well as two popular vertex-centric graph analytics system. 
Our results show that Vertica  has comparable end-to-end runtime performance, without requiring the use of a purpose-built graph engine.
In addition, we showed that using table UDFs, developers can trade memory footprint for reduced disk I/O and improve the performance of iterative queries even further.
In summary, the key takeaways from our analysis and evaluation in this paper are:

\noindent\textbf{Graph analytics can be expressed and tuned in RDBMSs.} Relational databases are general purpose data processing systems and graph operations can be expressed as relational operators in a relational database. Furthermore, relational databases can be effectively tuned to offer good performance on graph queries, yielding performance that is competitve with dedicated graph stores.

\noindent\textbf{Column-oriented databases excel at graph analytics.} We showed that column-oriented databases like Vertica can have comparable or better performance than popular graph processing systems (specifically, Giraph) on a variety of graph analytics. 
This is because these queries typically involves full scans, joins, and aggregates over large tables, for which Vertica is heavily optimized.
These optimizations include efficient physical data representation (columnar layout, compression, sorting, segmentation), pipelined query execution, and an optimizer to automatically pick the best physical plan.

\noindent\textbf{The RDBMS advantage: Graph + Relational Analysis.} Apart from outperforming graph processing systems on graph analysis, the real advantage of relational databases is the ability to combine graph analysis with relational analysis. This is because graph analysis is typically accompanied by relational analysis, either as a preparatory step or as a final reporting step. Such relational analysis is either impossible or highly inefficient in graph processing systems, while is natural to express in relational databases.

\noindent\textbf{Some graph queries are difficult to express in vertex-centric graph systems.} We showed that the vertex-centric programming model is a poor fit for several complex graph analyses. In particular, it is very difficult to program and run graph analysis involving 1-hop neighborhoods on vertex-centric systems.
This is because the scope of vertex-centric computations is restricted to a vertex and its immediate neighbors. In contrast, RDBMSs do not have such limitations and can efficiently execute 1-hop neighborhood computations.

\noindent\textbf{RDBMSs are a box full of goodies.} Relational databases come with many features that are not present (yet) in the next generation of graph processing systems. 
These include update handling, transactions, checkpointing and recovery, fault tolerance, durability, integrity constraints, type checking, ACID etc.
Current-generation graph processing systems simply do not come with this rich set of features.
One might think of stitching multiple systems together and coordinating between them to achieve these features, e.g.~using HBase for transactions and Giraph for analytics, but then we have the additional overhead of stitching these systems together. 
Of course, with time, graph-engines may evolve these features, but our results suggest that these graph systems should be layered on RDBMS, not built outside of them, which would enable them to inherit these features for free without giving up performance.
Furthermore, in many scenarios, the raw data for the graphs is maintained in a relational database in the first place. In such situations, it would make sense to be able to perform graph analytics within the same data management system.

\vspace{-0.4cm}

\bibliographystyle{IEEEtran}

\bibliography{references}

\vspace{-0.4cm}

\end{document}